\documentstyle[aps,amssymb,amsmath,graphicx,preprint]{revtex}
\tightenlines

\begin{document}

\draft

\title{On the collective behavior of parametric oscillators}
 
\author{I. Bena}
\address{Limburgs Universitair Centrum, B-3590 Diepenbeek, Belgium}
 
\author{R. Kawai}
\address{Department of Physics, University of Alabama at Birmingham,
Birmingham,  AL 35294}  

\author{C. Van den Broeck}
\address{Limburgs Universitair Centrum, B-3590 Diepenbeek, Belgium}
 
\author{Mauro Copelli\footnote{Present address: Instituto de
F\'{\i}sica, Universidade Federal Fluminense, Av. Litor\^anea, s/n - Boa
Viagem, 24210-340 Niter\'oi, RJ, Brazil} and Katja Lindenberg}
\address{Department of Chemistry and Biochemistry 0340 and Institute
for Nonlinear Science, University of California, San Diego,
La Jolla, CA 92093} 

\date{\today}
\maketitle
\begin{abstract}
We revisit the mean field model of globally and harmonically 
coupled parametric
oscillators subject to periodic block pulses with initially random phases. 
The phase diagram of regions of collective parametric
instability is presented, as is a detailed characterization 
of the motions underlying these instabilities.  This presentation
includes regimes not identified in earlier work [I. Bena and C. Van den
Broeck, Europhys. Lett. {\bf 48}, 498 (1999)].
In addition to the familiar parametric instability of individual
oscillators, two kinds of collective instabilities are identified.  In
one the mean amplitude diverges monotonically while in the other the
divergence is oscillatory.  The frequencies of collective oscillatory
instabilities in general bear no simple relation to the
eigenfrequencies of the individual oscillators nor to the frequency of
the external modulation.  Numerical simulations show that systems with
only nearest neighbor coupling have collective instabilities similar to
those of the mean field model.  
Many of the mean
field results are already apparent in a simple dimer [M. Copelli and K.
Lindenberg, to appear in Phys. Rev. E].
\end{abstract}


\section{Introduction}
\label{sec:intro}

Over the past two decades there has been an increasing interest in
the nonequilibrium behavior of spatially extended systems  
modeled as ensembles of simple dynamical units coupled to each other.
The collective evolution of such discrete coupled systems often
exhibits qualitatively different behavior from that of the
single units.  An example of a system that has attracted a great deal of
attention is a collection of a large number of coupled
limit-cycle (phase) oscillators with randomly distributed natural
frequencies~\cite{kuramoto}.  This system has been invoked as a
simple model for coupled chemical, biological or physical
oscillators.  A most spectacular collective phenomenon discovered
with this model is a synchronization phase transition involving
mutual entrainment of the oscillators through frequency and phase locking.
However, in this model the coupling is assumed to be sufficiently weak 
that the amplitude is not affected. As a result, the 
model cannot describe amplitude instabilities.  
A system that does exhibit a rich variety of amplitude instabilities
consists of coupled parametric oscillators and is
the subject of this paper~\cite{old,mauro}.  

There is a large literature on
single parametric oscillators with periodically~\cite{arnold} or
stochastically~\cite{linde} modulated frequencies, perhaps also
subject to thermal fluctuations~\cite{mazo}. A
wide variety of linear and nonlinear, deterministic and stochastic,
systems that exhibit energetic instabilities can be modeled as simple
parametric oscillators undergoing the amplitude instability known as
``parametric resonance"\cite{par}.
It is  therefore rather surprising that there is actually little work
on coupled parametric oscillators.
Work previous to our own that we are aware of is limited to
a global parametric modulation that acts on all the oscillators
in exactly the same way. Examples include parametrically pumped
electrons in a Penning trap~\cite{tan},
pattern formation under global resonant forcing~\cite{coullet}, 
time-periodic loading of an elastic system~\cite{nayfeh},
and nonlinear parametrically driven lattices~\cite{denardo}.

In a previous paper~\cite{old} we introduced a very simple,
analytically solvable model of coupled parametric oscillators
consisting of 
an infinite set of globally coupled harmonic oscillators subject to
time-periodic piecewise-constant modulations with randomly
distributed quenched phases (``quenched" in this context means that the
phase of the frequency modulation of each oscillator is set at time
$t=0$ and then remains unchanged).
We showed the appearance of a collective parametric instability:
even though each individual oscillator is in its stable parameter domain,
the average amplitude of the coupled system may diverge monotonically.
This collective instability occurs when the amplitude of parametric
modulation is sufficiently large that the instantaneous frequency of the
oscillators temporarily becomes imaginary.
The instability is re-entrant with respect to the strength of the
coupling of the oscillators and persists in the
overdamped limit. In the presence of a saturating
nonlinearity~\cite{kawai},
it generates a  pitchfork  bifurcation, corresponding
to a  genuine second-order nonequilibrium phase transition (implying the
spontaneous breaking of spatial symmetry and ergodicity).

The purpose of this paper is to show that {\em in addition} to the 
instability described by monotonic growth of the mean amplitude, 
the globally coupled infinite system can also
undergo transitions to a collective oscillatory instability
with an intrinsic frequency that is not connected in a simple 
way to either the frequencies of the 
individual oscillators or to the frequency of the external modulation.  
A saturating nonlinearity
turns this instability into a Hopf bifurcation generating
a limit-cycle with the concomitant breaking of 
temporal symmetry and ergodicity~\cite{kawai}.
Furthermore, we show that  even when the modulation amplitude is
very small, monotonic growth can still occur but via a
mechanism entirely different from that identified previously.

In trying to explain these collective 
behaviors, one is impressed by the 
similarities with a simple model introduced 
earlier~\cite{mauro} of {\em two} coupled parametric 
oscillators modulated periodically with a fixed phase difference 
$\theta$ between their modulations. 
The behavior of this dimer when 
$\theta = \pi$ is remarkably similar to that of the globally coupled
infinite system, and the roots of the instabilities observed 
in the latter are already present in this simple system.
At an even more primary level the seeds of these collective 
instabilities can be traced back to behaviors of
single oscillators.
In general an individual oscillator tends to synchronize with the
external 
modulation whereas the coupling induces mutual synchronization between
oscillators. These two synchronization tendencies cannot always be
satisfied simultaneously.
Coupling between oscillators can then be seen as leading to a sort of 
``selection'' among the single oscillator modes, enhancing some 
(destabilization) and smoothing out others (stabilization).

In Section~\ref{sec:model} we introduce the globally coupled model.
Section~\ref{sec:free} is a review of the properties of a single linear
parametric oscillator.  Section~\ref{sec:collective} establishes the
mathematical setting for the analysis of the globally coupled system as
a mean field problem, and typical numerical results are presented
in Section~\ref{sec:results}.  In Section~\ref{sec:phase} we collect
these results in the form of a phase diagram that characterizes the
behavior of the coupled system as the modulation parameters are varied.
Here we discuss the boundaries between stable and unstable
behavior and also
between different instability regimes.  Section~\ref{sec:dimer}
offers a
comparison between the mean field system and the
dimer.  Conclusions and a summary are presented in
Section~\ref{sec:conclusions}.

\section{The basic linear model}
\label{sec:model}

Consider a set of $N$ unit mass linear parametric oscillators with
displacements $\{x_i\}$, each with a periodically modulated frequency
and all harmonically 
coupled to one another.
Here we restrict our analysis to coupled {\em linear} parametric
oscillators.  The nonlinear case will be presented elsewhere~\cite{kawai}.
The equation of motion of the $i$-th oscillator
is given by
\begin{equation}
\ddot{x}_i + \gamma \dot{x}_i + \omega_0^2 [1+ \xi_i(t)] 
x_i = - \frac{k}{N} \sum _{j=1}^{N} (x_i - x_j)\,\,, 
\label{eq:EOM}
\end{equation} 
with $i=1,\cdots,N$.  

Analytic results are possible
with a simple piecewise-constant periodic modulation (period=$T_p$),
defined as
\begin{equation}
\xi_i(t) = A \;\text{sgn}[ \sin \omega_p (t+\tau_i) ]
\label{eq:perturbation}
\end{equation}
where $\omega_{p}=2 \pi / T_p$, 
and the initial phase $\tau_{i}$ is chosen at random for each
oscillator.  We are mainly interested in the mean amplitude
\begin{equation}
\left< x \right> = \frac{1}{N} \sum_{i=1}^N x_i
\label{eq:average1}
\end{equation}
as a measure of the macroscopic behavior of the system. In the 
thermodynamic limit $N\rightarrow\infty$, this site average 
(\ref{eq:average1}) is equivalent to the average with respect to the 
random phase $\tau_{i}$ of the displacement of a single oscillator $i$,
\begin{equation}
\left< x \right> = \frac{1}{T_{p}} \int_0^{T_{p}} x_{i} \; 
d\tau_{i}\,\,,
\label{eq:average2}
\end{equation} 
which is independent of $i$. Equation~(\ref{eq:EOM}) 
can then be reduced to a single mean-field differential equation, 
\begin{equation}
\ddot{x} + \gamma
\dot{x} + \omega_0^2[1+\xi_\tau(t)]x = 
-k(x-\left< x \right> )
\label{eq:EOM_mf}
\end{equation} 
where we have dropped the index $i$ and have now indexed the modulation 
with the initial phase chosen at random for each oscillator. Note 
that the average $\left< x \right>$ must be evaluated self-consistently 
using Eqs.~(\ref{eq:average2}) and (\ref{eq:EOM_mf}).

\section{Review of The Single Parametric Oscillator}
\label{sec:free}

The further analysis of the coupled system is clarified by first
briefly examining the properties of a single parametric oscillator
subject to the piecewise-constant frequency modulation
(\ref{eq:perturbation})~\cite{arnold,nayfeh,vanderpol,yorke,linde_book,preprint}, for
which the equation of motion is simply
\begin{equation}
\ddot{x} + \gamma \dot{x} + \omega_0 [1 + \xi_\tau (t)]x = 0.
\label{eq:EOM_free}
\end{equation}
The arbitrary phase $\tau$ can be set to zero, but is retained 
explicitly 
in order
to make clearer the connection with the coupled system.
The equation of motion
can be solved using Floquet theory (see the Appendix),
and the result may be written as
\begin{equation}
\left ( \begin{array}{c}
x(t) \\ \dot{x}(t)+\frac{\displaystyle \gamma}
{\displaystyle 2}x(t)
\end{array} \right )
= {\sf G}_\tau (t)
\left ( \begin{array}{c}
x(0) \\ \dot{x}(0)+\frac{\displaystyle \gamma}
{\displaystyle 2}x(0)
\end{array} \right )
\end{equation}
where ${\sf G}_\tau(t)$ is the time-evolution operator,
whose explicit expression
is given in the
Appendix.

Floquet theory provides information about the state of the
system at the end of each modulation period, but it does not  
determine the time evolution {\em within} each period.  
If, as in most of the literature, stability conditions were
our only concern, 
finding the Floquet eigenvalues would be sufficient.  However, we are
interested not only in the conditions giving rise to 
instabilities but also in the types of instabilities. 
Since exponential decay or growth of the oscillator amplitude
is expected in any case, the time evolution may be conveniently 
investigated by Laplace transformation,
\begin{eqnarray}
\tilde{x}(s) &=& \int_0^\infty e^{-st} x(t) dt \nonumber \\ [12pt]
&=& [\tilde {\sf G}_\tau(s)]_{11} x(0)+[\tilde{\sf G}_\tau(s)]_{12} \left (
\frac{\gamma}{2} x(0) + \dot{x}(0)\right ),
\label{eq:Laplace}
\end{eqnarray}
where $[\tilde{\sf G}_\tau(s)]_{ij}$ is a matrix element of the
Laplace transform of the time-evolution operator. 
In general, if $\tilde{x}(s)$ has poles at 
$s_{j} = \Lambda_j + i \Omega_j$, 
the temporal behavior of the amplitude is expressed as
\begin{equation}
x(t) = \sum_j c_j e^{\Lambda_j t} e^{i\Omega_j t}
\label{eq:amplitude}
\end{equation}
where the
$c_j$'s are constants determined by the inverse Laplace
transform and the initial conditions, 
\begin{equation}
c_j=2\pi i \lim_{s\rightarrow s_j} (s-s_j) \left\{ \left[ \tilde{\sf G}_\tau
(s)\right]_{11} x(0) +\left[ \tilde{\sf G}_\tau (s)\right]_{12} \left(
\frac{\gamma}{2} x(0) +\dot{x}(0) \right) \right\}.
\end{equation}The Laplace transform of the 
time-evolution operator is derived in
the Appendix
as 
\begin{equation}
\tilde{\sf G}_\tau(s) =
\frac{\frac{1}{2} e^{-\gamma T_{p}/2}
\left[e^{(s+\gamma)T_{p}} I  - {\sf G}^{-1}_\tau(T_{p}) \right] }
{\cosh \left[\left(s+\frac{\displaystyle \gamma}{\displaystyle
2}\right)T_{p}\right]- R }  
\int_0^{T_{p}} e^{-st} {\sf G}_\tau(t) dt\,\,,
\label{eq:Gs} 
\end{equation}
where $I$ is the identity matrix and the resonance
parameter $R$ is defined as
\widetext
\begin{equation}
R = \cos\left(\frac{\omega_{+}T_{p}}{2}\right) 
    \cos\left(\frac{\omega_{-}T_{p}}{2}\right) 
    -\frac{\omega_{+}^{2} + \omega_{-}^{2}}{2\omega_{+} \omega_{-}} 
    \sin\left(\frac{\omega_{+}T_{p}}{2}\right)  
    \sin\left(\frac{\omega_{-}T_{p}}{2}\right) 
\label{eq:R}
\end{equation}
with
\begin{equation}
\omega_\pm = \omega_0 \sqrt{ 1 \pm A -
\left( \frac{\gamma}{2\omega_0}\right)^2 }.
\label{eq:omega_pm}
\end{equation}
The poles of Eq.~(\ref{eq:Gs}) are determined by the condition
\begin{equation}
\cosh\left[\left(s+\frac{\gamma}{2}\right)T_{p}\right]-R = 0\,\,,
\label{eq:Free_Singularity1}
\end{equation}
and their real parts are explicitly given by
\begin{equation}
\Lambda_j = \left\{
\begin{array}{l l}
-\frac{\gamma}{2} \pm \frac{1}{T_{p}}\ln \left[ -R + \sqrt{R^2
- 1} \right] & \text{for} \qquad R\leq -1 \\
-\frac{\gamma}{2}  & \text{for} \qquad  |R| \le 1 \\
-\frac{\gamma}{2} \pm \frac{1}{T_{p}}\ln \left[ R + \sqrt{R^2
- 1} \right] & \text{for} \qquad R \geq 1\;,
\end{array}
\right.
\label{eq:Lambdaj}
\end{equation}
while the imaginary parts are
\begin{eqnarray}
& & \frac{\Omega_j}{\omega_p}  = \left\{
\begin{array}{l l}
j-\frac{1}{2} & \text{for} \; R\leq -1  \\ 
j & \text{for} \; R \geq 1 \;,
\end{array}
\right. \nonumber \\
& & \frac{\Omega_{j}^{\pm}}{\omega_p}
= j \pm \frac{1}{2 \pi} \arccos(R) 
\qquad \text{for}\; |R| \leq 1\;.
\label{eq:Omegaj}
\end{eqnarray}
Here $\arccos(R)$ lies in the range $[0,\,\pi]$,
$j$ is an integer.
Note that the real parts of the poles are in fact
independent of $j$, and therefore one  can  drop the subscript $j$ on 
$\Lambda$ and rewrite the amplitude (\ref{eq:amplitude}) as
\begin{equation}
x(t) = e^{\Lambda t} \sum_j c_j e^{i\Omega_j t}.
\label{eq:sum}
\end{equation}
Correspondingly, the relative weights of the different modes
remain the same for all times: they all die ($\Lambda < 0$), or they all 
explode ($\Lambda > 0$), or they all keep their initial amplitudes 
($\Lambda =0$). None becomes relatively more dominant with time.

\begin{figure}
\begin{center}
\includegraphics[width=2.8in]{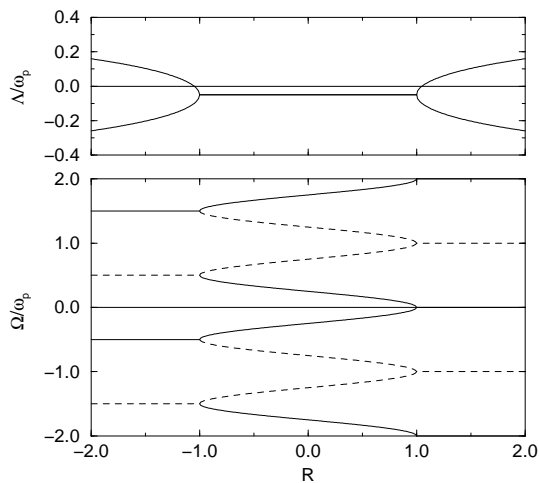}
\caption{Real (upper graph) and imaginary (lower graph) parts of the poles
$s_j$ as a
function of $R$ with  $\gamma/\omega_p=0.1$.  The behavior associated
with these poles is described in detail in the text.  Note the
significance of the boundaries $R=\pm 1$, and of
the boundaries $R=\pm R_c$ where one $\Lambda$ becomes positive.
The latter mark the onset of single-oscillator parametric instability.}
\label{fig:poles}
\end{center}
\end{figure}

Figure \ref{fig:poles} illustrates the dependence of
the poles on the resonance parameter.  
Consider first the real part $\Lambda$.  When $|R|<1$ it is
single-valued, $R$-independent, and negative
(unless there is no damping $\gamma$, in which case
it vanishes and the motion is purely oscillatory). 
At the bifurcation points $R=\pm 1$ the real
parts become $R$-dependent but remain negative until
$|R|$ reaches the value
\begin{equation}
R_c \equiv \cosh\left ( \frac{\gamma T_{p}}{2}\right ) 
\label{eq:Rc}
\end{equation}            
which is greater than unity unless $\gamma=0$.  Beyond $|R|=R_c$
one of the $\Lambda$'s becomes positive, leading to
exponential growth of the oscillator amplitude.
The condition $|R| = R_c$ thus
corresponds to the 
onset of parametric resonance or, more appropriately,
of parametric instability. 

\begin{figure}
\begin{center}
\includegraphics[width=2.8in]{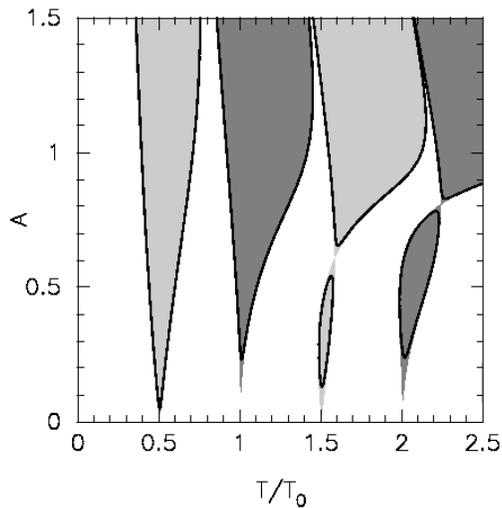}
\caption{Shaded regions indicate $R>1$ (darker) and $R<-1$ (lighter).
Solid lines delineate the boundaries $|R|=R_c$.  The damping
$\gamma=0.01$ and the frequency $\omega_0=0.4$.}
\label{fig:R}
\end{center}
\end{figure}

Consider next  the imaginary parts $\Omega_j$, shown in 
Fig.~\ref{fig:poles} for $j=-2,\ldots 2$ alternately 
as solid and dashed curves.  The condition $|R|=1$ 
here also marks an interesting boundary
at which the behavior of the imaginary parts changes from being
$R$-independent when $|R|>1$ to $R$-dependent when $|R|<1$.
On the large-$|R|$ side of this boundary the frequencies
$\Omega_j$ are simply proportional to the frequency
$\omega_p$ of the modulation. When $|R|<1$, however, the oscillator
frequencies change continuously with 
$R$ and bear no simple
relation to either $\omega_p$ or the natural frequency
$\omega_0$. 
In Fig.~\ref{fig:R} we present the regions in parameter space $(A, \, 
T_{p}/T_{0})$ where $|R| >1$. The darker regions correspond to $R>1$ and 
the lighter to $R<-1$. The stability boundaries $|R|=R_{c}$ are also 
indicated by solid lines, the oscillator being unstable inside these 
boundaries. Note that the $|R|=R_{c}$ and $|R|=1$ frontiers almost 
coincide because of the very low damping ($\gamma 
=0.1$). As is well known, for small $A$ the instability 
appears in the vicinity of $T_{p}\approx j T_{0}/2$ ($j$ integer).

Although the boundary $|R|=R_{c}$ is important in determining the 
transition from stable to unstable behavior for the single parametric 
oscillator, it does not play the same role for the coupled system, 
for which the boundary $|R|=1$ turns out to acquire further 
significance. Indeed, as we will see in the following section, each 
individual pole $s_{j}=\Lambda + i \Omega_{j}$ gives rise
to a collective pole in the coupled system, and these collective poles 
have different 
$\Lambda$'s. Therefore, some modes may become unstable ($\Lambda >0$),
and therefore 
dominant, even when $|R|<1$, while others remain stable ($\Lambda <0$).
A very striking example is that of the single oscillator mode with
a zero imaginary 
part, i. e., the one with $j=0$ when $R>1$. If $\Lambda$ is negative, it 
simply provides a monotonically decaying contribution to the oscillator 
displacement, and if $\Lambda$ is positive it provides a 
monotonically growing contribution. However, its relative weight 
in the sum (\ref{eq:sum}) is extremely small and therefore this mode 
is practically not detectable in simulations of a single 
oscillator.  Its contribution is overwhelmed by those of the
oscillatory modes.
However, in the coupled system this non-oscillatory mode may become 
dominant because it may become unstable while the 
oscillatory modes are still stable. This will lead to 
a monotonic explosion of the mean.

\section{Collective Instabilities}
\label{sec:collective}

The mean field equation (\ref{eq:EOM_mf}) can be rearranged as
\begin{equation}
\ddot{x} + \gamma \dot{x} + \{\omega_0^2[1+\xi_\tau(t)] +k\}x =
k \left< x \right>
\label{eq:EOM_mf2}
\end{equation}
which describes a single parametric oscillator of frequency
\begin{equation}
\omega_k = \omega_0 \sqrt{1+\frac{k}{\omega_0^2}}
\label{eq:omega_k}
\end{equation}
driven by an ``additive force'' $k\left< x \right>$.
Therefore, it can be solved with the time-evolution operator
(\ref{eq:GT}) of a single oscillator except that Eq.~(\ref{eq:omega_pm})
must be replaced with the new shifted frequencies

\begin{equation}
\omega_\pm = \omega_0 \sqrt{ 1 \pm A  +\frac{k}{\omega_0^2} -
\left ( \frac{\gamma}{2\omega_0}\right )^2 } .
\label{eq:omega_pm2}
\end{equation}

The general solution of Eq.~(\ref{eq:EOM_mf2}) can be written as

\begin{equation}
\left( 
      \begin{array}{c}
         x(t)\\
         \dot{x}(t) + \frac{\gamma}{2}x(t) 
      \end{array}
\right)
= {\sf G}_\tau (t)
\left(
      \begin{array}{c}
         x(0)\\
         \dot{x}(0) + \frac{\gamma}{2}x(0)
      \end{array}
\right)
+ \ k {\sf G}_\tau(t) \int_0 ^{t}
{\sf G}_\tau(t')^{-1} 
\left(
      \begin{array}{c}
         0\\
         \left< x(t')\right>
      \end{array}
\right) dt' .
\label{eq:integral_EOM}
\end{equation} 
Taking the average of Eq.~(\ref{eq:integral_EOM}) with respect to the
random phase, we obtain a self-consistent equation for 
the mean amplitude,
\begin{equation}
\left< x(t)\right> =  \left<[{\sf G}_\tau(t)]_{11}\right>x(0)
+ \left<[{\sf G}_\tau(t)]_{12}\right>
\left (
\frac{\gamma}{2}x(0) + \dot{x}(0) \right )
+ k \int_0 ^{t} K(t-t') \left<x(t')\right> dt',
\label{eq:SCF}
\end{equation}
where the kernel is defined by
\begin{equation}
K(t-t') = \left<\left[{\sf G}_\tau(t) {\sf G}_\tau^{-1}(t')\right]_{12}
\right> \,\,.
\label{eq:kernel}
\end{equation} 
The kernel (\ref{eq:kernel})
depends only on the time difference because of the {\em uniform} 
distribution of the initial phases $\tau$. One can therefore solve 
the integral equation (\ref{eq:SCF}) 
by Laplace transformation.  The solution in Laplace space is given by

\begin{equation}
\left<\tilde{x}(s)\right> = \frac{
\left<[{\tilde G}_\tau(s)]_{11}\right>x(0)
+\left<[{\tilde G}_\tau(s)]_{12}\right> \left (
\frac{\gamma}{2} x(0) + \dot{x}(0) \right )}
{1 - k\left<[\tilde{G}_\tau(s)]_{12}\right>}\,\,.
\label{eq:xs}
\end{equation}
Equation~(\ref{eq:xs}) has a set of poles determined by the condition
\begin{equation}
\left<[\tilde{\sf G}_\tau(s)]_{12}\right> = \frac{1}{k}\,\,,
\label{eq:pole2}
\end{equation}
which differ from the ones derived for a single parametric oscillator.
When $k \rightarrow 0$ the right hand side of Eq.~(\ref{eq:pole2})
diverges and the poles of the coupled system clearly approach 
those of the single oscillator. 
By explicitly performing the time integral in Eq.~(\ref{eq:Gs}) and
taking the average over the random phase $\tau$, one can obtain a rather 
complicated but explicit expression of Eq.~(\ref{eq:pole2}) as
\begin{eqnarray}
&T_{p}& \omega_{+} \omega_{-} 
\left [
      \cosh\left(\frac{\tilde{\gamma}T_{p}}{2}\right) - R 
\right ] \nonumber\\ [12pt]
& &\times \:
\left( \omega_{+}^2 + \frac{\tilde{\gamma}^2}{4}\right)^2
\left( \omega_{-}^2 + \frac{\tilde{\gamma}^2}{4}\right)^2 
\left [
      \frac{
              k \left(\omega_{+}^2 + \omega_{-}^2 +
                \frac{ \displaystyle \tilde{\gamma}^2} 
                {\displaystyle 2} \right)}
           { 2 \left(\omega_{+}^2 +
                \frac {\displaystyle \tilde{\gamma}^2} {\displaystyle 4}\right)
               \left(\omega_{-}^2 + \frac{\displaystyle \tilde{\gamma}^2}
                       {\displaystyle{4}}\right) } 
       - 1 
 \right ]
\nonumber\\ [12pt]
& &-\; k \omega_{+} (\omega_{+}^2 - \omega_{-}^2)^2 
                        \left(\omega_{-}^2 - \frac{\tilde{\gamma}^2}{4}\right) 
  \sin\left(\frac{\omega_{-}T_{p}}{2}\right) 
  \left[ \cosh \left(\frac{\tilde{\gamma}T_{p}}{4}\right) -
    \cos\left(\frac{\omega_{+}T_{p}}{2}\right) \right] \nonumber\\ [12pt]
& &-\; k \omega_{-} (\omega_{+}^2 - \omega_{-}^2)^2 
                        \left(\omega_{+}^2 - \frac{\tilde{\gamma}^2}{4}\right)
  \sin\left(\frac{\omega_{+}T_{p}}{2}\right) 
  \left[ \cosh \left(\frac{\tilde{\gamma}T_{p}}{4}\right) - 
   \cos\left(\frac{\omega_{-}T_{p}}{2}\right) \right] \nonumber\\ [12pt]
& &-\; k \omega_{+} \omega_{-} \tilde{\gamma}
  (\omega_{+}^2 - \omega_{-}^2)^2 \sinh \left(\frac{\tilde{\gamma} 
  T_{p}}{4}\right)
  \left[ 2 \cosh\left(\frac{\tilde{\gamma}T_{p}}{4}\right) -
    \cos\left(\frac{\omega_{+}T_{p}}{2}\right) -
    \cos\left(\frac{\omega_{-}T_{p}}{2}\right)\right]
\nonumber\\ [12pt]
& &=\; 0
\label{eq:pole4}
\end{eqnarray}
where
\begin{equation}
\tilde{\gamma} \equiv \gamma + 2 s = \gamma + 2\Lambda + 2i\Omega.
\label{eq:gamma2}
\end{equation}
Finding the complex roots of Eq.~(\ref{eq:pole4}) is difficult even
numerically. Therefore, we first investigate the collective modes
graphically. Graphical inspection not only provides qualitative
understanding of collective modes but also helps to identify suitable
numerical algorithms. 

Coupling between the oscillators causes a frequency shift of each
oscillator as indicated in Eq.~ (\ref{eq:omega_pm2}).
This is a trivial effect that is accounted for by calculating all
single oscillator quantities such as $R$ and the poles
$\Lambda + i\Omega$ using the {\em shifted} frequency.  The
interesting questions concern the way in which the
couplings between the shifted oscillators affect the dynamics, and,
in particular, whether instabilities and synchronization may be caused
or suppressed by the coupling.  A great deal can be learned about
these dynamics starting from the single oscillator roots shown in
Fig.~\ref{fig:poles}.  We thus investigate the coupled
system in three typical regimes: (1) A single oscillator with the
given (shifted) parameters
is unstable ($R\le -R_c$);  (2) The single oscillator is
stable and in the regime where its frequency is determined
by the modulation frequency ($1 \le R \le R_c$); (3) The
single oscillator is stable and its frequency bears no
simple relation to the natural (shifted) oscillator frequency nor to the
modulation frequency ($|R| \le 1$).  The results are presented in
the next section.

\section{Results}
\label{sec:results}

To find the poles associated with collective modes we need to solve
the coupled equations
\begin{eqnarray}
& &k\text{Re}\left<[\tilde{\sf G}_\tau(s)]_{12}\right>-1 = 0 
\label{eq:Real}\\ [12pt]
& &\text{Im}\left<[\tilde{\sf G}_\tau(s)]_{12}\right> = 0
\label{eq:Imag}
\end{eqnarray}
for $\Lambda$ and $\Omega$ in $s=\Lambda +i\Omega$.  In the next set of
figures we present results for the three cases listed at the end of
the last section.
Each case is presented in two or three figures.  

In one set of
figures we plot
the left-hand sides of Eqs.~(\ref{eq:Real}) 
and (\ref{eq:Imag}) 
as contour lines in the space 
($\Lambda/\omega_p$, $\Omega/\omega_p$) [the solid lines for 
Eq.~(\ref{eq:Real}), and the broken ones for Eq.~(\ref{eq:Imag})].  
The parameters $k=1.28$, $\omega_0=0.4$,
$\gamma=0.01$, and
$A=1.0$ are used in all the figures. 
The thick lines indicate zero contours. Therefore,
the solutions we seek are 
the intersections of these two sets of thick lines, 
indicated by open
circles. The poles of single 
oscillators  with shifted frequency are indicated
by black solid circles.
There is of course an infinite number of solutions but we
only exhibit those that fall within the scale of our figures.
  Associated with each case is also a figure
representing a number of trajectories as a function of time
that characterize the particular
situation.  For some of the examples we also show the associated phase
space trajectories.  Note that in this presentation we make a careful
distinction between a {\em single} oscillator and an {\em individual}
oscillator.  The former refers to an independent
oscillator with parameters
$\omega_k$ and $\gamma$, while the latter refers to one of the oscillators
in a coupled system with parameters 
$\omega_k$, 
$\gamma$, and $k$.  

\begin{figure}
\begin{center}
\includegraphics[width=2.8in]{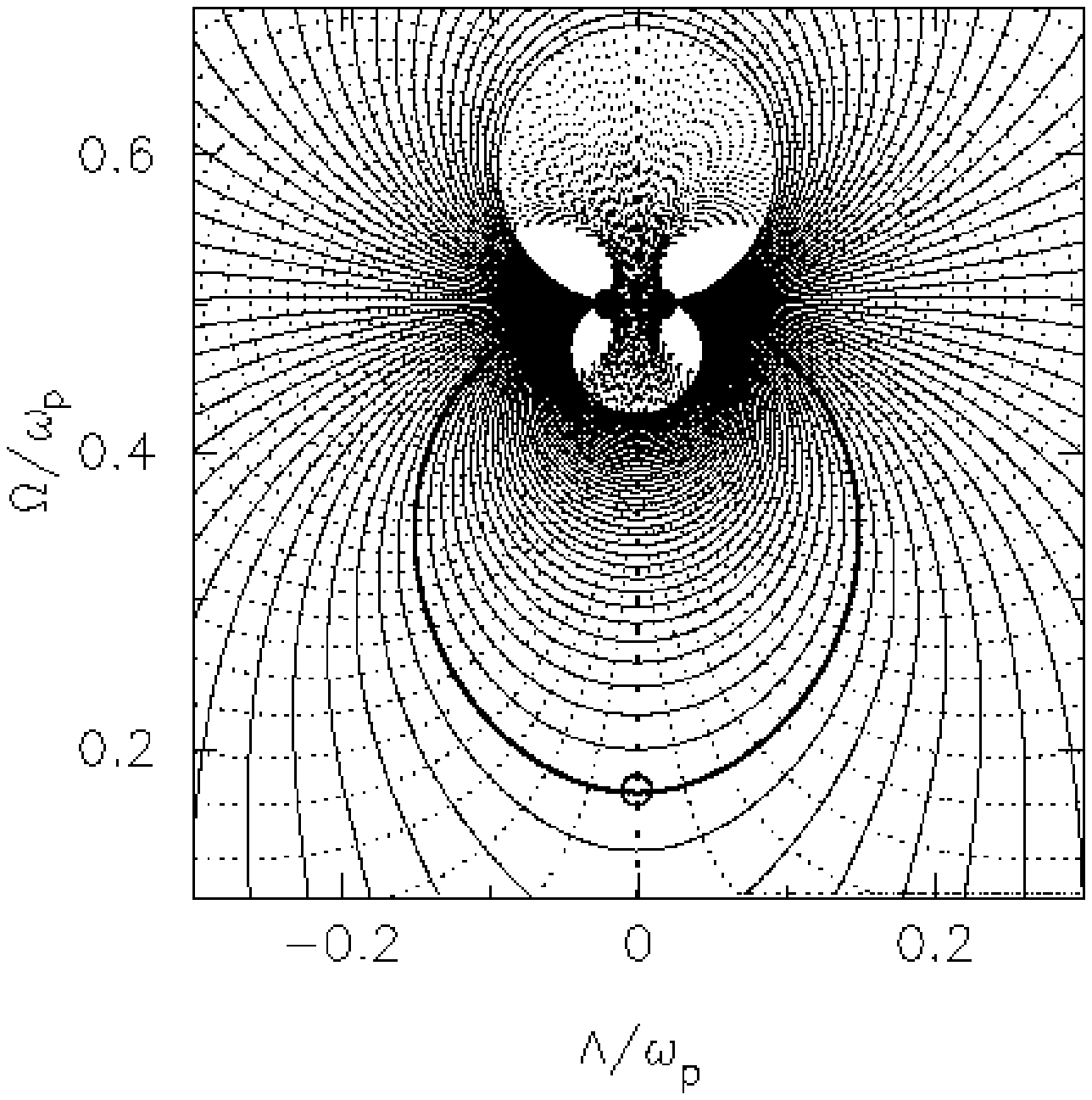}
\caption{Contour plots of the left-hand side of Eq.~(\ref{eq:Real})
(solid lines) and Eq.~(\ref{eq:Imag}) (broken lines) for  
$\omega_0=0.4$, $k=1.28$, $\gamma=0.01$, $A=1.0$, and $T_k=2T_{p}$,
which lead to $R=1.00617$.  
Poles of a single oscillator are indicated by solid circles.
Thick lines correspond to solutions of Eqs.~(\ref{eq:Real})
(solid) and (\ref{eq:Imag}) (broken). The intersections of the
solid and dotted thick lines, indicated by open circles, are thus 
solutions of Eq.~(\ref{eq:pole2}). Although there is a single oscillator
pole with positive $\Lambda$ there is no positive-$\Lambda$ collective
mode.}    
\label{fig:mode1}
\end{center}
\end{figure}

Figure \ref{fig:mode1} shows the contours and poles for case (1).  Here
$\omega_k=\omega_p/2$,
which corresponds to $R=-1.0062$.  Since $R<-1$, the poles of
single oscillators appear on
the lines $\Omega_j/\omega_p=(j-\frac{1}{2})$ each as a pair because
in this regime there are two values of $\Lambda$ associated with each
$\Omega$.  Only the poles for $j=1$ are shown because the others (and
there are an infinite number of them) are off the scale of the figure. Since
for these parameters $|R|>R_c=1.00077$, one of each pair of poles has
a positive $\Lambda$. The amplitude of
a single oscillator would therefore diverge exponentially. 
However, as the numerical calculations indicate, all
the collective modes 
have $\Lambda=-\gamma/2$ and thus the
mean amplitude decays to zero despite the instability of 
single oscillators.

\begin{figure}
\begin{center}
\includegraphics[width=3.8in]{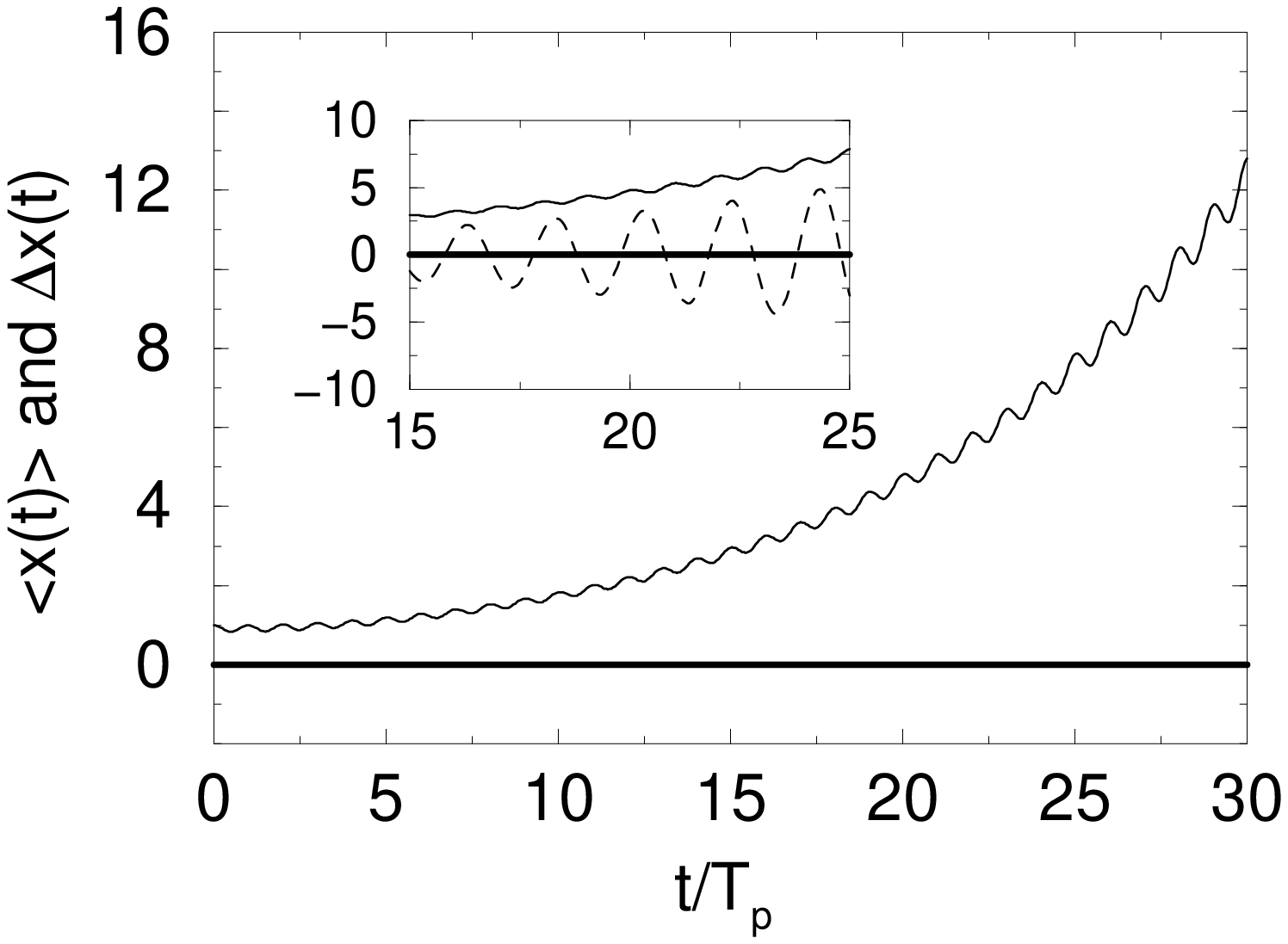}
\caption{Trajectories associated with Fig.~\ref{fig:mode1}.  Thick solid
line: mean $\left< x \right>$.  Thin solid line: 
deviation $\Delta x$.  The inset also includes the trajectory $x$ of an 
individual oscillator (broken line).} 
\label{fig:sim1}
\end{center}
\end{figure}

\begin{figure}
\begin{center}
\includegraphics[width=4.8in]{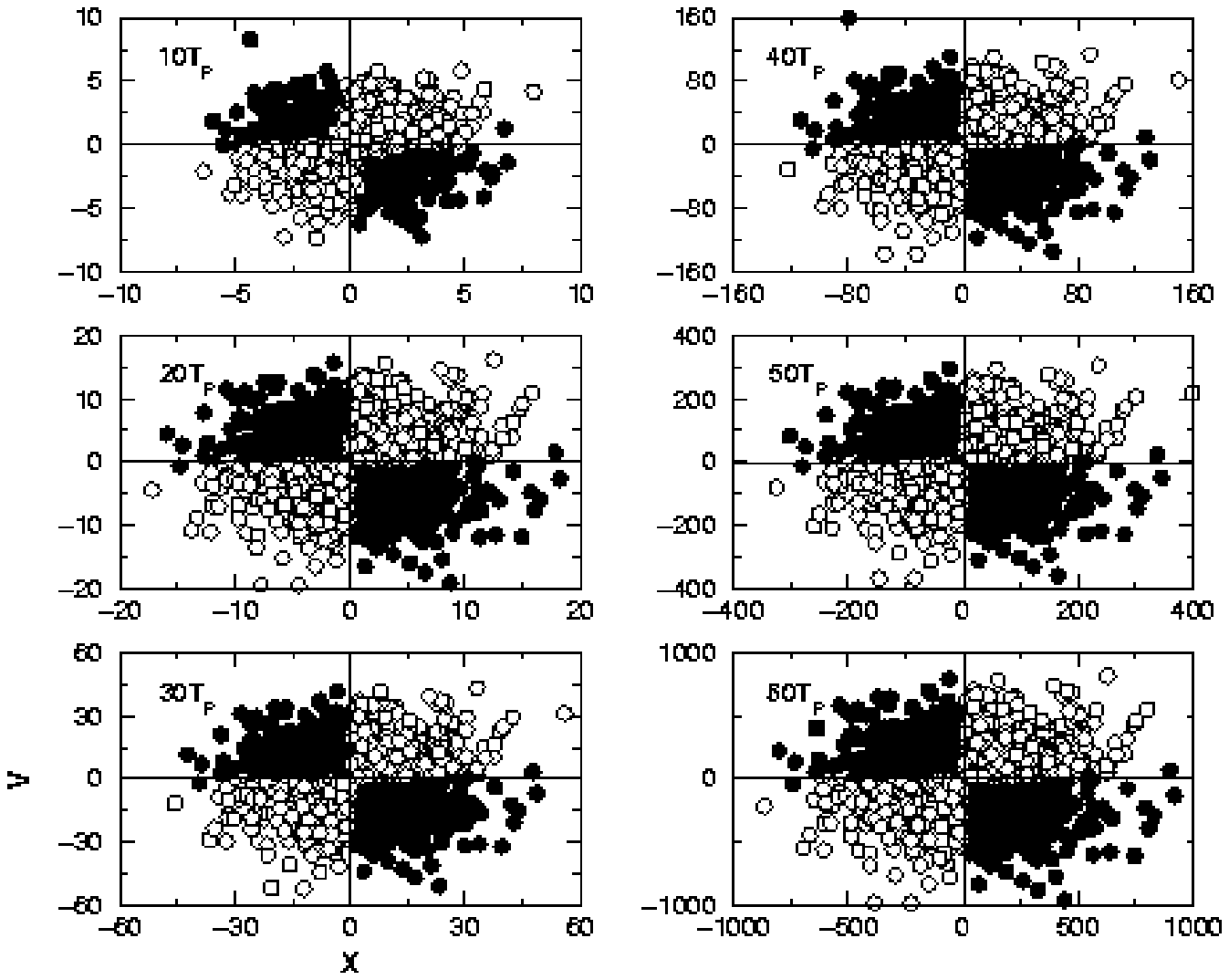}
\caption{Phase point snapshots of $2000$ individual oscillators in the
system associated with Figs.~\ref{fig:mode1} and \ref{fig:sim1}.
Note the scale changes with increasing time.}
\label{fig:phstrj1}
\end{center}
\end{figure}

Computer simulation results for various trajectories of a system
of 100,000 oscillators with these parameters are shown in 
Fig.~\ref{fig:sim1}.   We see that the mean amplitude is indeed zero,
and that the deviation $\Delta x = \sqrt{\left<x^2\right>-
\left<x\right>^2}$ diverges.  The inset shows the same two trajectories
as well as the diverging trajectory of an individual oscillator.
Phase space
trajectories of individual oscillators in the coupled system are
shown in Fig.~\ref{fig:phstrj1}. Each circle indicates a snapshot of
an individual oscillator.  Solid circles represent oscillators with
positive modulation and open circles are ones with negative modulation at 
the time of the snapshot.  Only 1000 oscillators out of 100,000
are shown.

The six snapshots show
that with increasing time the phase volume increases (note the different
scales in each snapshot), which is consistent with
the divergent behavior of each oscillator and with the growth of the
deviation $\Delta x$, and provide confirmation
that there is indeed no mutual synchronization nor other kinds of
organized collective motion.  However, the persistent separation of 
solid and open circles into separate quadrants indicates
that individual oscillators
are synchronized with the external modulation.  Note that any {\em
individual} oscillator moves clockwise, switching colors accordingly.
However, since the phase
of the modulation is chosen at random for each oscillator, 
at any time half of the oscillators have positive modulation and the
other half negative, so that the pattern in preserved and an average
over the phase gives $\left<x\right>=0$. 
In this case, the effect of the coupling is only the frequency shift
(\ref{eq:omega_pm2}).  When the coupling strength is increased at
fixed $T_p/T_k$, mutual synchronization will eventually be established.  
Then, the individual oscillators must lose synchronization with the
modulation and 
consequently parametric instability is suppressed.
Therefore, the coupling
stabilizes the system in this parameter region.

\begin{figure}
\begin{center}
\includegraphics[width=2.8in]{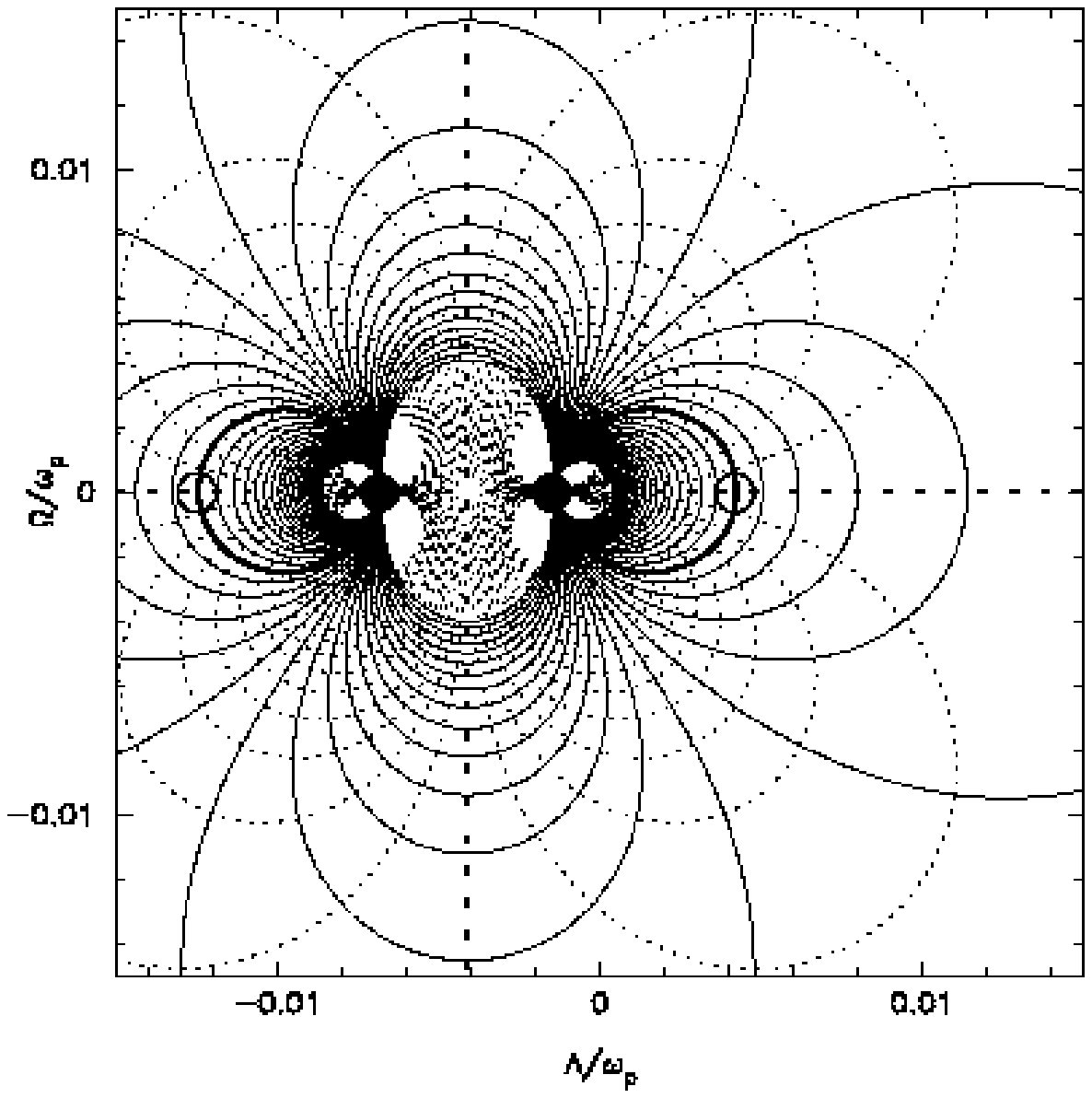}
\caption{Contour plots of the left-hand side of Eqs.~(\ref{eq:Real})
(solid lines) and Eq.~(\ref{eq:Imag}) (broken lines) for  
$\omega_0=0.4$, $k=1.28$, $\gamma=0.01$, $A=1.0$, and $T_k=T_{p}$, 
which lead to $R=1.0001$.  
Poles of a single oscillator are indicated by solid circles.
Thick lines correspond to solutions of Eqs.~(\ref{eq:Real}) (solid) and
(\ref{eq:Imag}) (broken).  The intersections of the solid and dotted
thick lines, indicated by open circles, are thus solutions of
Eq.~(\ref{eq:pole2}).  Although the single oscillator poles  have all
negative $\Lambda$, the collective modes include a pole
with positive $\Lambda$ and $\Omega=0$.  This mode diverges
exponentially without oscillation.}
\label{fig:mode2}
\end{center}
\end{figure}

Interesting behavior is observed when $\omega_k=\omega_p$, that is,
the single oscillator frequency is equal to the modulation frequency.
This corresponds to case (2) on our list. The contours and poles for
this case are shown in Fig.~\ref{fig:mode2}.  Since $R=1.0001$, the 
single oscillator has poles at $\Omega_j=j\omega_p$. Only the poles for
$j=0$ are shown.  This condition is close to parametric resonance,
but since $R$ is just below $R_c=1.0031$ all the single oscillator modes
have negative $\Lambda$'s and therefore single oscillator trajectories
decay to the absorbing state $x=0$.
However, one of the collective modes has a pole with $\Omega=0$
and a {\em positive} $\Lambda$.  All the other collective modes
are dominated by this $j=0$ unstable mode and therefore
the mean amplitude diverges monotonically. 

\begin{figure}
\begin{center}
\includegraphics[width=3.8in]{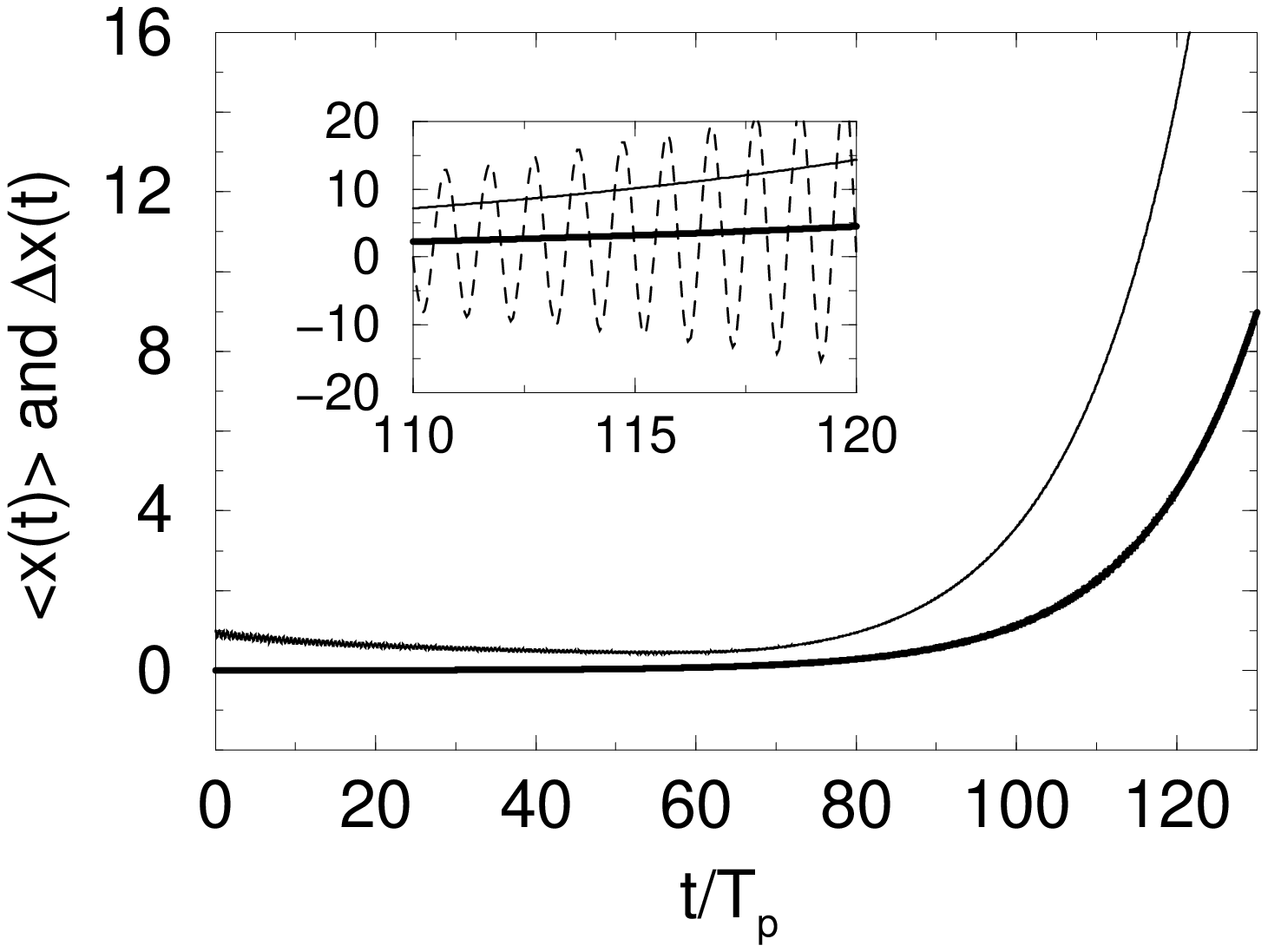}
\caption{Trajectories associated with Fig.~\ref{fig:mode2}.  Thick solid
line: mean $\left< x \right>$.  Thin solid line:
deviation $\Delta x$.  The inset also includes the trajectory $x$ of an
individual oscillator (broken line).}
\label{fig:sim2}
\end{center}
\end{figure}

\begin{figure}
\begin{center}
\includegraphics[width=4.8in]{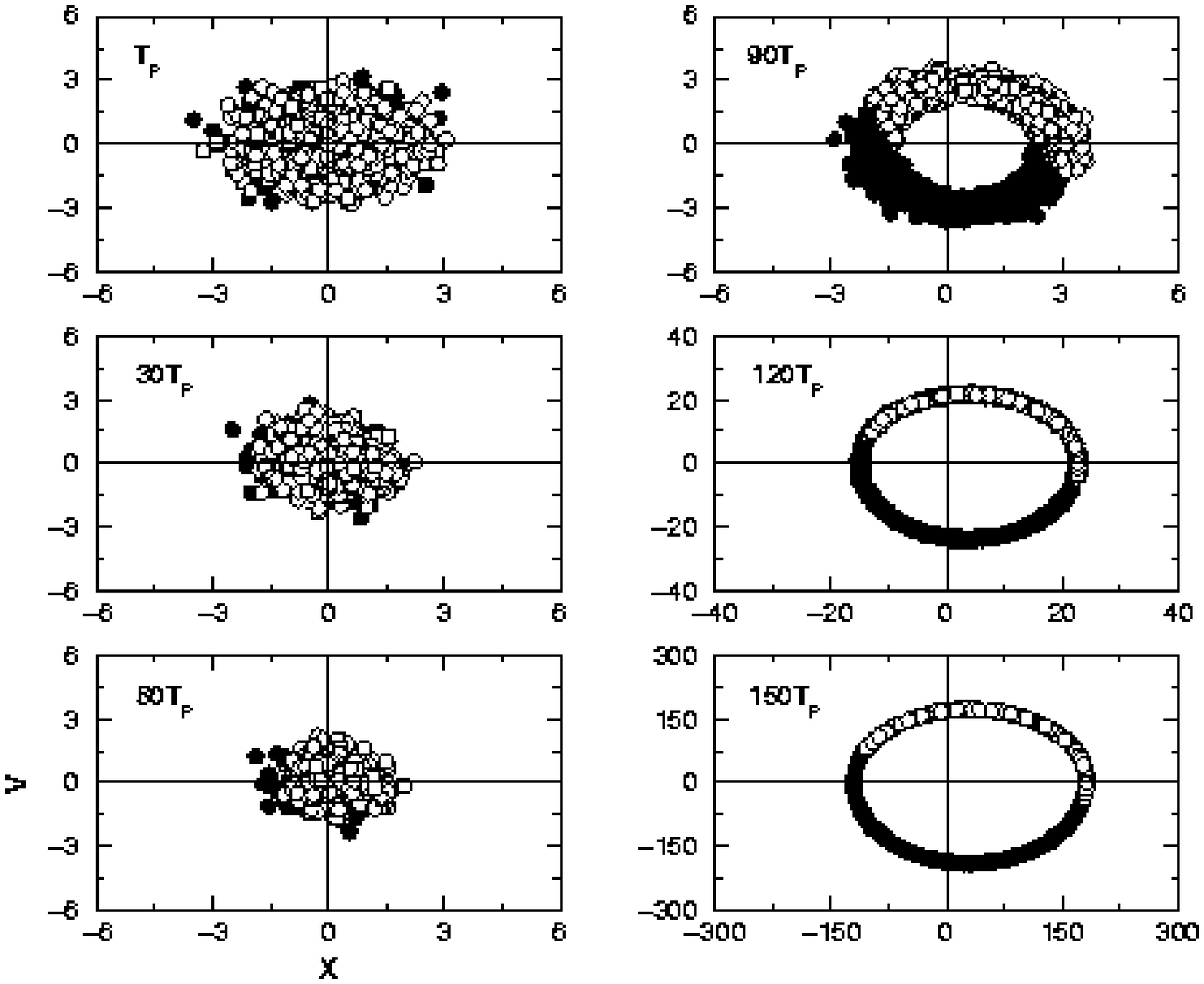}
\caption{Phase point snapshots of $2000$ individual oscillators in the
system associated with Figs.~\ref{fig:mode2} and \ref{fig:sim2}.
Note the scale changes with increasing time.}
\label{fig:phstrj2}
\end{center}
\end{figure}

Computer simulation results for the associated trajectories are
shown in Figure \ref{fig:sim2}.  
The mean decays slowly at the beginning and then diverges 
monotonically.  The deviation diverges as well, and does so more
rapidly.
The individual oscillator trajectory also diverges; in this
example, although each single oscillator would be stable, the coupling
causes individual oscillators in the system to become unstable.
In other words, each oscillator is driven by the
diverging mean on the right hand side of Eq.~(\ref{eq:EOM_mf2}).
The phase space trajectories 
in Fig.~\ref{fig:phstrj2} show that after an initial transient (first
three panels, where open circles are hiding most of the solid circles),
individual oscillators in the coupled system oscillate
with increasing amplitude about $\left<x\right>$, while the mean 
$\left<x\right>$ is moving away from the origin. In the long time 
limit, each oscillator ``forgets'' its initial conditions and it is 
driven by the mean. Therefore, while the phase of each individual 
oscillator is determined by the phase of the modulation, the 
amplitude in phase space is determined by the mean. 
Correspondingly, the oscillators 
become ``amplitude-synchronized'' through the mean. 
Until the synchronization is well established, the individual
oscillators decay because the single oscillator modes have negative
$\Lambda$.

As in the previous case, the
phases of all single-oscillator modes with $\Omega_j \ne 0$ 
become synchronized with
the external modulation.  The separation of solid and open circles in
the three later-time panels in
Fig.~\ref{fig:phstrj2} indicates this synchronization.
However, in contrast with the previous case, 
there is now a zero-frequency mode which does not have a phase
to be synchronized.  This mode is therefore not affected by
the phase of each oscillator nor by the phase of the modulation. 
The zero-frequency mode shifts the center of oscillation 
away from $x=0$ in either the positive or the negative direction. 
In the presence of coupling the oscillators tend to follow the mean, 
and therefore shift in the
same direction, thus breaking the system symmetry. 
Note that the open and solid circles no longer lie entirely in separate
quadrants.  It is precisely the excess of open circles relative to solid
ones in the positive quadrants (and vice versa in the negative
quadrants) that drives the motion of the mean further to the right.
This mode can now dominate the behavior of the system because,
contrary to the situation of  
a single oscillator, each mode has a different $\Lambda$. 
In this example the
effect of coupling is thus not only to shift the frequency according to
(\ref{eq:omega_pm2}), but also the more interesting collective
symmetry-breaking monotonic divergence of the mean amplitude and the
mutual synchronization of the individual oscillator amplitudes.

Since the role of the coupling in this case is  
to suppress the single-oscillator modes with $\Omega_j \ne 0$ 
and to allow the mode with $\Omega_j=0$ to grow, 
this instability persists even for very large $k$.  
It should be noted that this collective instability
is a different phenomenon from the monotonic growth reported 
previously\cite{old}.  That instability occurs only for 
large $A$ and disappears when the coupling strength $k$ is increased
(re-entrant transition).

\begin{figure}
\begin{center}
\includegraphics[width=2.8in]{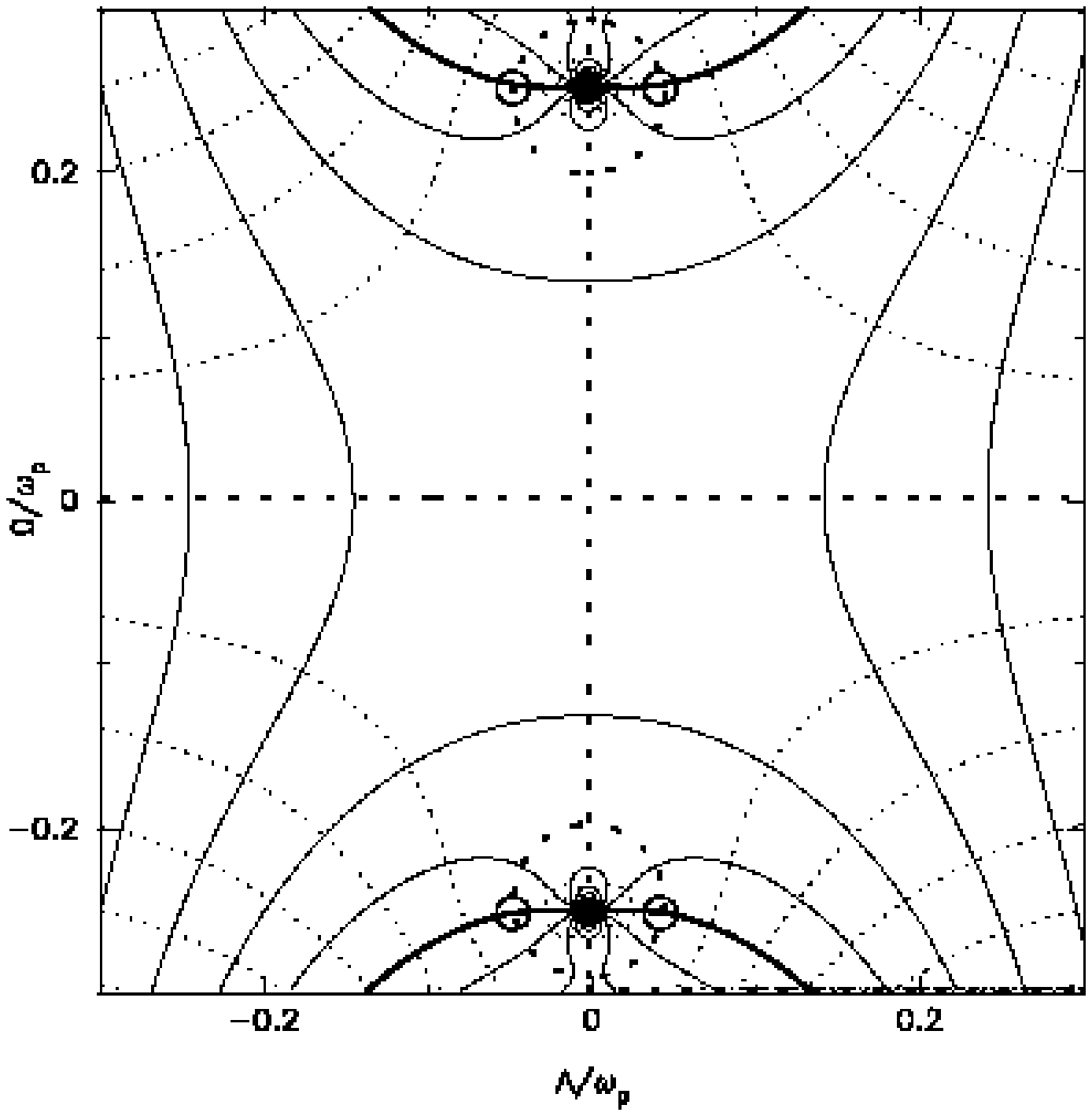}
\caption{Contour plots of the left-hand side of Eqs.~(\ref{eq:Real})
(solid lines) and Eq.~(\ref{eq:Imag}) (broken lines) for
$\omega_0=0.4$, $k=1.28$, $\gamma=0.01$, $A=1.0$, and $T_k=4T_{p}/3$,
which lead to $R=-0.01037$.
Poles of a single oscillator are indicated by solid circles.
Thick lines correspond to solutions of Eqs.~(\ref{eq:Real}) (solid) and
(\ref{eq:Imag}) (broken).  The intersections of the solid and dotted
thick lines, indicated by open circles, are thus solutions of
Eq.~(\ref{eq:pole2}).  Although the single oscillator poles all have
negative $\Lambda$, the collective modes include a pole
with positive $\Lambda$ and nonzero.  This mode diverges
exponentially with oscillation.}         
\label{fig:mode3}
\end{center}
\end{figure}

\begin{figure}
\begin{center}
\includegraphics[width=3.8in]{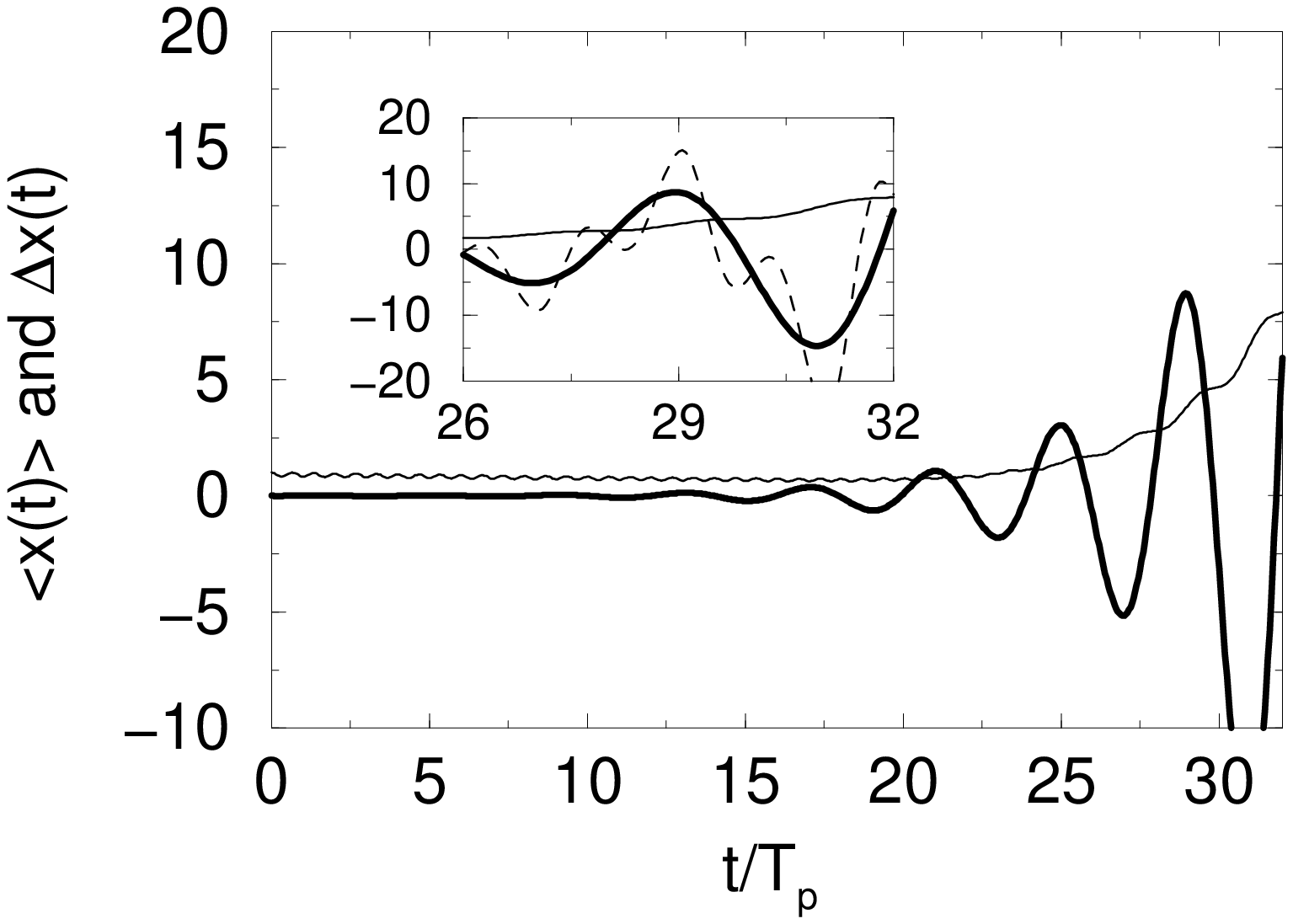}
\caption{Trajectories associated with Fig.~\ref{fig:mode3}.  Thick solid
line: mean $\left< x \right>$.  Thin solid line:
deviation $\Delta x$.  The inset also includes the trajectory $x$ of an
individual oscillator (broken line).}    
\label{fig:sim3}
\end{center}
\end{figure}

\begin{figure}
\begin{center}
\includegraphics[width=4.8in]{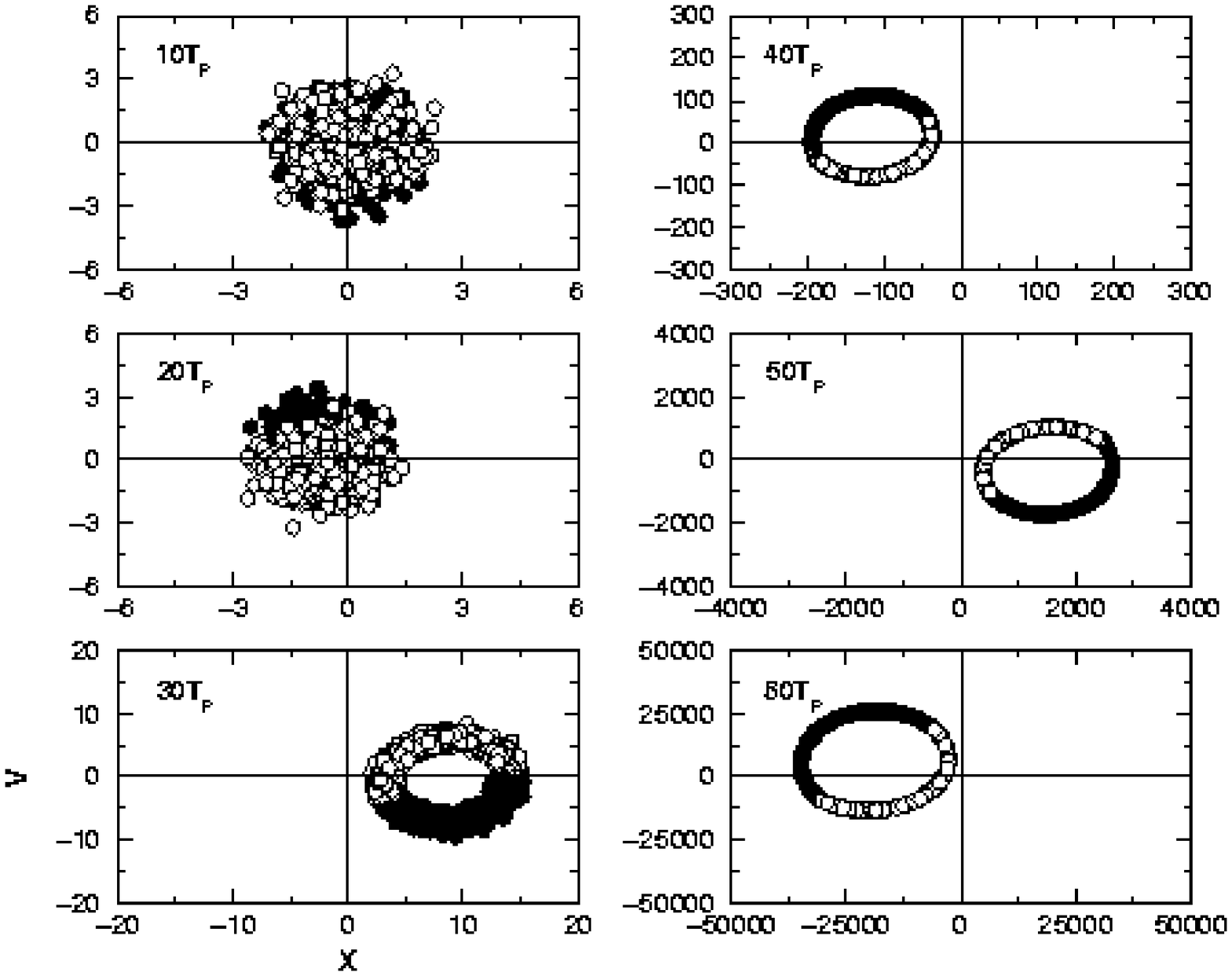}
\caption{Phase point snapshots of $2000$ individual oscillators in the
system associated with Figs.~\ref{fig:mode3} and \ref{fig:sim3}.
Note the scale changes with increasing time.}
\label{fig:phstrj3}
\end{center}
\end{figure}

Finally, case (3) occurs when $\omega_{k}$ 
is between the two previous
cases, so that $|R|<1$ and single oscillators are not in parametric
resonance condition.  However,
Fig.~\ref{fig:mode3} indicates that at least one of the collective modes 
has a positive $\Lambda$ as well as non-zero $\Omega$.
Therefore, the mean $\left<x\right>$ oscillates with diverging 
amplitude.
Recall that when $|R|<1$ the eigenfrequencies  of single
oscillators vary continuously with $R$ 
but do not match the frequency
of the modulation nor the natural frequency of the oscillator. 
Therefore, the individual synchronization to the external
modulation plays no role and the phases of the oscillators are free to
synchronize to one another
through a synchronization to the phase of the mean. 
Computer simulations shown in Fig.~\ref{fig:sim3} confirm the
oscillatory instability, and the phase space points of individual
oscillators shown in Fig.~\ref{fig:phstrj3} testify to this phase 
synchronization. Although solid and open circles again form separate
groups, the entire ring of open and solid circles alternates between the
positive quadrants and the negative quadrants.
Therefore, all oscillators are mutually synchronized.

In the next section we collect the results for the coupled system into
a phase diagram indicating regions of stability and instability of
different types.  

\section{Phase Diagram}
\label{sec:phase}

A convenient way to summarize observations and characterize the
instabilities systematically is in the form of appropriate 
``phase diagrams" in which the stability boundaries are presented
as a function of the system parameters.  Since there are many parameters
in this model the full diagram would involve a many-dimensional
representation.  Instead, we present the diagram in the two-dimensional
space $(A, T_{p}/T_k)$ that characterizes the external
modulation for a given set of oscillator parameters
$\omega_k$, $k$, and $\gamma$. 

\begin{figure}
\begin{center}
\includegraphics[width=4.8in]{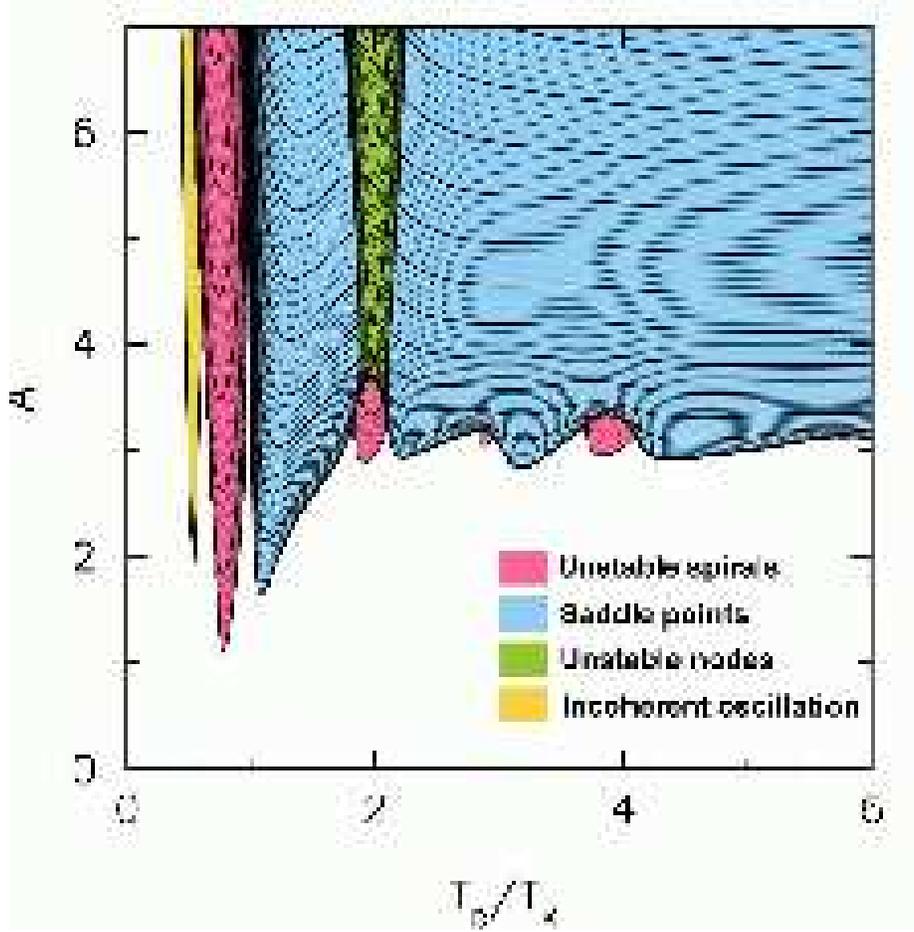}
\vspace{0.3in}  
\caption{
Phase diagram for the mean field model with oscillator parameters
$\omega_0=0.4$, $k=1.28$ and $\gamma=0.16$.  White regions denote stable
regimes.  The various instability regimes are color coded as indicated.
The characteristic behavior in each instability regime is described in
the text.}
\label{fig:phasehighgamma}
\end{center}
\end{figure}

Figure~\ref{fig:phasehighgamma} shows the phase diagram for the system
parameters indicated in the caption.  The colored regions denote
unstable regimes, each color coding a particular type of instability.
The yellow region denotes parameter ranges where the individual
oscillators are unstable but the mean amplitude for the coupled system
is zero (``incoherent unstable oscillations").  This region is
associated with divergences of the {\em first term} in
Eq.~(\ref{eq:integral_EOM}) but {\em not} of the second term.  In this
case the distinction between ``single oscillators" and ``individual
oscillators" becomes moot, since the right hand side of
Eq.~(\ref{eq:EOM_mf2}) plays no role.  Oscillatory instabilities of
the mean with positive $\Lambda$ and nonzero $\Omega$ are denoted in
pink (``unstable spirals").  Blue regions (`` saddle nodes") denote
monotonic divergence of the mean with one positive $\Lambda$ and zero
$\Omega$.  Green regions (``unstable nodes") also denote monotonic
instabilities but with two positive $\Lambda$'s and zero $\Omega$.
These latter three instabilities involve divergences of the {\em
second term} in Eq.~(\ref{eq:integral_EOM}).  The first term may
diverge as well, i.e., the $\Lambda$'s of the single oscillator and of
the collective mode may be simultaneously positive. When this is the
case we always observe the $\Lambda$ of the collective mode to be
larger than that of the single oscillator modes; the collective mode
therefore dominates the dynamics.  In establishing the terminology for
various instabilities we have loosely followed the usual conventions
of nonlinear dynamics.

We can associate each specific case discussed in the previous section
with a location on this sort of phase diagram.  Thus the trajectories
in Fig.~\ref{fig:sim1} are in a yellow regime of incoherent
oscillatory divergence of each oscillator.  Those of
Fig.~\ref{fig:sim2} are in a blue or green regime of monotonic
divergence of the mean, and those of Fig.~\ref{fig:sim3} are in the
pink regime of unstable spirals.

\begin{figure}
\begin{center}
\includegraphics[width=4.8in]{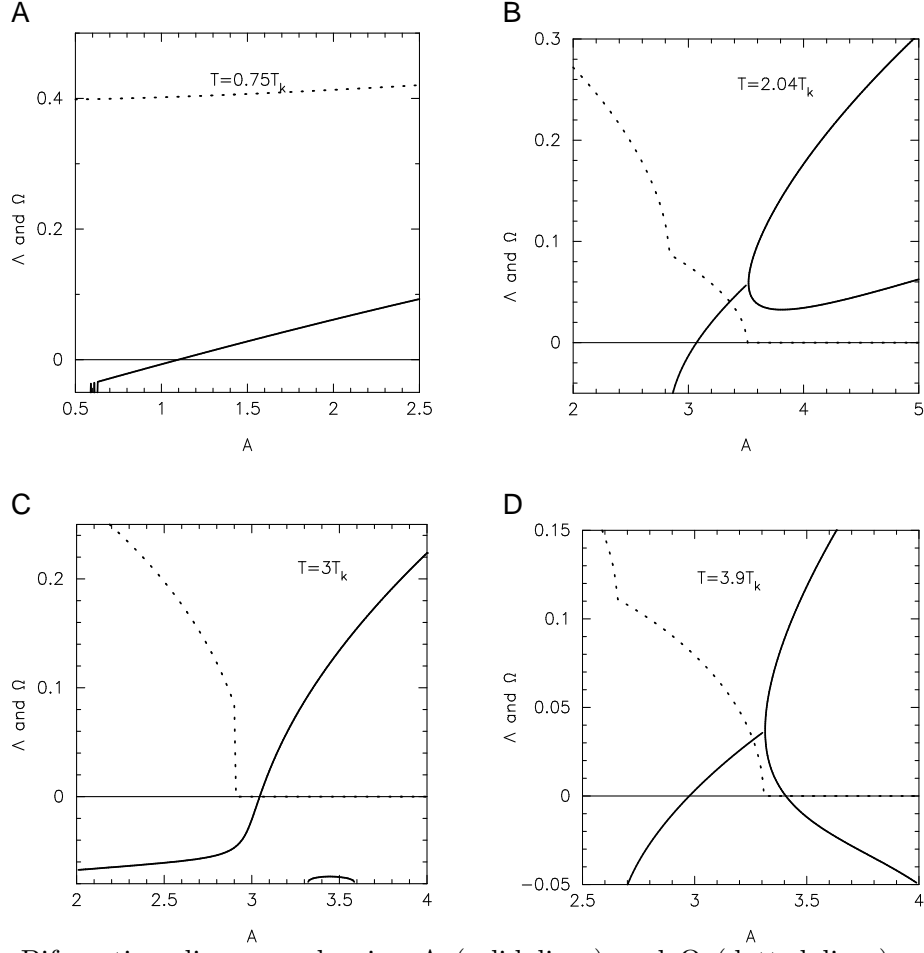}
\caption{Bifurcation diagrams showing
$\Lambda$ (solid lines) and $\Omega$ (dotted lines) with changing
modulation amplitude for various values of the modulation period.
Panel (a): $T_{p}/T_k=0.75$; (b): $T_{p}/T_k = 2.04$; (c):
$T_{p}/T_k=3.0$; (d): $T_{p}/T_k=3.9$. The behavior implied by these
diagrams is discussed in detail in the text.}
\label{fig:bifurcation}
\end{center}
\end{figure}

It is helpful to follow the behavior of the oscillator system across
the various collective instability boundaries by considering the signs
of $\Omega$ and $\Lambda$ as one increases the modulation amplitude
$A$ (thus moving upward vertically along the phase diagram) for
different fixed values of the modulation period $T_{p}$.  Various
associated bifurcation diagrams presenting $\Lambda$ (solid lines) and
$\Omega$ (dotted lines) as a function of $A$ are shown in
Fig.~\ref{fig:bifurcation}.  Consider first the period $T_{p}=0.75
T_k$, shown in panel (a).  As $A$ increases, $\Lambda$ changes sign,
becoming positive at $A_c=1.14$ while $\Omega$ remains positive.  This
represents a transition from a stable spiral to an unstable spiral
(pink region in Fig. \ref{fig:phasehighgamma}).

Consider next the period  $T_{p}=2.04T_k$, shown in panel (b).
Here $\Lambda$ becomes positive at
$A_c=3.06$ while $\Omega$ remains positive. This therefore again marks
a transition from a stable spiral to an unstable spiral (pink region).
However, with a further increase in amplitude, $\Omega$ eventually
goes to zero at $A_c^\prime=3.51$, where $\Lambda$ bifurcates to two
positive values (an unstable node, green region) via a saddle-node
bifurcation.  The oscillatory instability thus switches to a monotonic
instability at this point.

A different transition pattern is seen when $T_{p}=3.0T_k$, shown in
panel (c).  It begins with a stable spiral and switches to a stable
node at $A_c=2.91$. With a further increase in amplitude the system
undergoes a transition to a saddle node (blue region) at
$A_c^\prime=3.04$.

A more complex transition pattern is shown in panel (d), in which the
character of the instability changes several times along the
$T_{p}=3.9T_k$ line. As usual, at low amplitudes there is a stable
spiral.  At the point $A_c=2.99$ the system moves into an unstable
spiral (pink region).  The unstable spiral becomes an unstable node
(very small green region in the phase diagram) via a saddle-node
bifurcation at $A_c^\prime=3.30$.  A further transition to a saddle
node (blue) occurs at $A_c^{\prime\prime} = 3.4$.

\begin{figure}
\begin{center}
\includegraphics[width=4.8in]{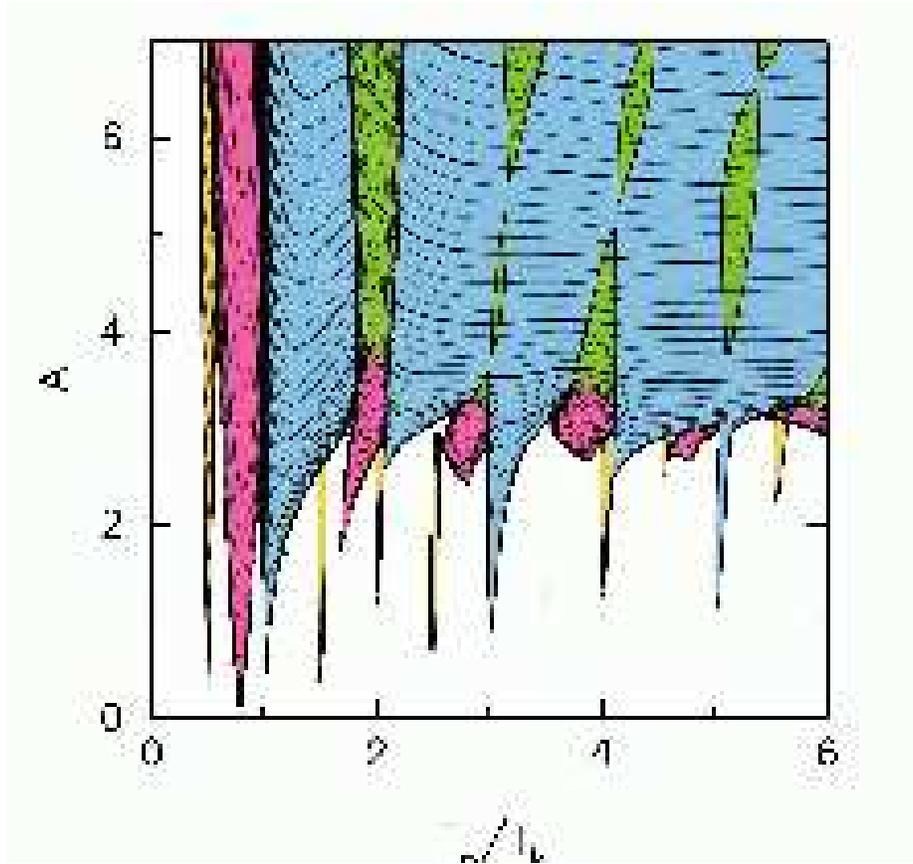}
\vspace{0.3in}  
\caption{ Phase diagram for the mean field model with oscillator parameters
$\omega_0=0.4$, $k=1.28$ and $\gamma=0.01$.}  
\label{fig:phaselowgamma}
\end{center}
\end{figure}

We note that the phase diagram just described is quite rich and
intricate.  Instabilities cover even larger regions in parameter
space, and do so with increasing intricacy, as the damping $\gamma$
decreases.  A typical phase diagram for low damping is shown in
Fig.~\ref{fig:phaselowgamma}.

Increasing coupling or damping also leads to greater stability: the
stability boundaries move ``up" in the phase diagram when $\gamma$
and/or $k$ are increased, indicating that a stronger modulation is
needed to cause unstable behavior.  Furthermore, the oscillatory
instabilities eventually disappear with increasing modulation period
$T_p$, leaving only the monotonic collective instabilities.  However,
it should be noted that the monotonic instabilities for sufficiently
large modulation period are simply due to an inversion of the
effective harmonic potential and hence not due to any collective
effects.  That there is an inversion can already be suspected from the
fact that at least one of the shifted frequencies in
Eq.~(\ref{eq:omega_pm2}) becomes imaginary.

An analysis of the system for large $T_p$ is fairly simple and
instructive in elucidating the source of instabilities more
explicitly.  In the adiabatic limit $T_p\rightarrow\infty$ the single
oscillator frequencies are frozen in time, half of them at frequency
$\tilde{\omega}_+$ and the other half at frequency $\tilde{\omega}_-$,
where
\begin{equation}
\tilde{\omega}_\pm = \omega_0 \sqrt{1+\frac{k}{2\omega_0^2} \pm A}.
\label{eq:dimer_omega}
\end{equation}        
The mean field equations of motion then are
\begin{eqnarray}
\left<\ddot{x}\right>_+ + \gamma \left<\dot{x}\right>_+
+ \tilde{\omega}_+^2 \left<x\right>_+ &=&
\frac{k}{2}\left<x\right>_- \nonumber \\ [12pt]
\left<\ddot{x}\right>_- + \gamma \left<\dot{x}\right>_-
+ \tilde{\omega}_-^2 \left<x\right>_- &=&
\frac{k}{2}\left<x\right>_+
\label{eq:dimer}
\end{eqnarray}       
where $\left<\cdots\right>_\pm$ indicates an average over the
oscillators with the frequency $\tilde{\omega}_{\pm}$ respectively.
This system can be diagonalized analytically in full generality. The
eigenmodes of the coupled system are characterized by the
complex frequencies
\begin{eqnarray}
\Omega^{(1)}_\pm & = & i\frac{\gamma}{2} \pm \omega_0
\sqrt{ 1 + \frac{k}{2\omega_0^2} - \frac{\gamma^2}{4\omega_0^2} 
+ \sqrt{A^2+\left(\frac{k}{2\omega_0^2}\right)^2} } \nonumber \\
\Omega^{(2)}_\pm & = & i\frac{\gamma}{2} \pm \omega_0
\sqrt{ 1 + \frac{k}{2\omega_0^2} - \frac{\gamma^2}{4\omega_0^2} 
- \sqrt{A^2+\left(\frac{k}{2\omega_0^2}\right)^2} } .
\label{eq:eigenfrequencies}
\end{eqnarray}        
This case clearly illustrates the distinction between what we have
called ``single oscillator instabilities" and ``chain instabilities".
The former refer to the frequency (\ref{eq:dimer_omega}) while the
latter refer to (\ref{eq:eigenfrequencies}). While the single
oscillators would remain stable until $A = 1+k/2\omega_0^2$ (at which
point $\tilde{\omega}_-$ becomes imaginary), the chain becomes
destabilized when $A$ reaches the value $\sqrt{1+k/\omega_0^2}$, where
the imaginary part of $\Omega^{(2)}_-$ becomes negative (note that the
transition point is independent of $\gamma$).  Beyond that the system
is in a saddle-point/unstable-node instability region of
non-oscillatory exponential 
growth due to potential inversion, which was the primary reason of
instability in the previous work\cite{old}.  In the phase diagrams
Figs.~\ref{fig:phasehighgamma} and \ref{fig:phaselowgamma}
this translates to a stability boundary that settles at $A=3$
as $T_p\rightarrow \infty$.  Note that in particular the boundary
$A=\sqrt{1+k/\omega_0^2}$ remains valid in the overdamped limit
$\gamma\to\infty$, consistent with our early work\cite{kawai98}.

The origin of the instabilities presented as narrow
blue and pink tongues in the low-$A$ region of 
Fig.~\ref{fig:phaselowgamma} is entirely different from the 
mechanism based on the temporarily inverted potential and unique
to the underdamped case.  In the following sections we will
explain the cause of 
these instabilities using a dimer of parametric oscillators.

\section{Comparison With Parametric Oscillator Dimer}
\label{sec:dimer}

In a recent paper~\cite{mauro} we reported results for a model of two
coupled oscillators subject to parametric modulations with a phase
difference $\theta$.  The equations of motion for this system are
just the $N=2$ version of Eq.~(\ref{eq:EOM}):
\begin{eqnarray}
\label{eq:twoosc}
\ddot{x}_{1} & = & -\omega_{0}^2[1+\xi_{1}(t)]x_{1} - \frac{k}{2}(x_{1}-x_{2})
-\gamma \dot{x}_1
\nonumber \\
\ddot{x}_{2} & = & -\omega_{0}^2[1+\xi_{2}(t)]x_{2} - \frac{k}{2}(x_{2}-x_{1})
-\gamma \dot{x}_2 .
\end{eqnarray} 
The piecewise-constant periodic modulations of the two oscillators
differ by a constant phase $\theta$, so that we can write
Eq.~(\ref{eq:perturbation}) for this case as
\begin{eqnarray}
\label{eq:phasedif}
\xi_{1}(t) & = & A\;\mbox{sgn}\left(\sin(\omega_{p}t)\right)
\nonumber \\[12pt]
\xi_{2}(t) & = & A\;\mbox{sgn}\left(\sin(\omega_{p}t+\theta)\right) .
\end{eqnarray}         
We want to
investigate whether the mean position $x \equiv (x_1 + x_2)/2$ reproduces
the macroscopic behavior of the mean in the globally coupled model.

In the absence of parametric modulations, a dimer has two eigenmodes:
symmetric (or {\it mutually synchronized\/}, $x_1(t) = x_2(t)$) and
antisymmetric (or {\it mutually antisynchronized\/}, $x_1(t) =
-x_2(t)$), with the former having a lower energy.
When the parametric modulations are applied, these modes are in general
no longer the eigenmodes of the dimer (except for $\theta=0$).  However,
the motions can be expressed as linear combinations of these modes and,
in particular, the behavior of the mean of interest is reflected in
the excitation of the symmetric mode by the parametric modulations.

In previous work~\cite{mauro} we focused on the behavior of the system
as a function of the parameters $A$, $T_p=2\pi/\omega_p$, $\omega_0$,
$\gamma$, $k$, and $\theta$.  Among our conclusions is the fact that
the regions of parametric instability are sensitively dependent on the
phase difference $\theta$. Of particular interest for the analysis in
this paper is the behavior of the anti-phased dimer
($\theta=\pi$). This particular dimer captures many of the features of
the mean field coupled system with unexpected accuracy. This
assertion, which was originally based on our previous comparison of
the regions of parametric resonance~\cite{mauro}, is reinforced when
the dimer bifurcation diagrams are further refined to take into
account details of qualitatively different trajectories, as we shall
see below.

For the piecewise constant parametric modulation~(\ref{eq:phasedif})
the solution of the dimer problem is formally simple. The stability
analysis is based on the eigenvalues $\{\lambda_{j}\}$
($|\lambda_1|\ge |\lambda_2|\ge |\lambda_3|\ge |\lambda_4|$) of the
Floquet operator~\cite{mauro}: parametric instability occurs when
$|\lambda_{1}|>1$. To characterize different types of parametric
instability we present bifurcation diagrams using the same color
conventions as in the phase diagrams of the mean field model
(Figs.~\ref{fig:phasehighgamma} and~\ref{fig:phaselowgamma}).  If
$\mbox{Im } \lambda_{1}\neq 0$, then clearly one has an oscillatory
instability (pink regions). If $\mbox{Im }\lambda_{1}=0$, the
instability can be either oscillatory or monotonic. Using the
eigenvector corresponding to $\lambda_{1}$ as initial condition, we
have determined whether or not $x$ crosses zero during one period of
the modulation. If it does, the point is assigned to a pink region. If
it does not, the second largest eigenvalue $\lambda_{2}$ will
determine whether the point belongs to a blue ($|\lambda_{2}|\le 1$)
or green ($|\lambda_{2}|> 1$) region. The pink, blue and green regions
are all caused by the instability of the symmetric mode. The yellow
region, on the other hand, requires the instability of the
antisymmetric mode {\it and\/} the decay of the symmetric
mode. However, such a purely antisynchronous solution is forbidden in
the $\theta=\pi$ case~\cite{mauro}, which means that yellow regions
cannot appear at all in the anti-phased dimer~\cite{phasedimer}.

\begin{figure}[!bt]
\begin{center}
\includegraphics[height=4.8in]{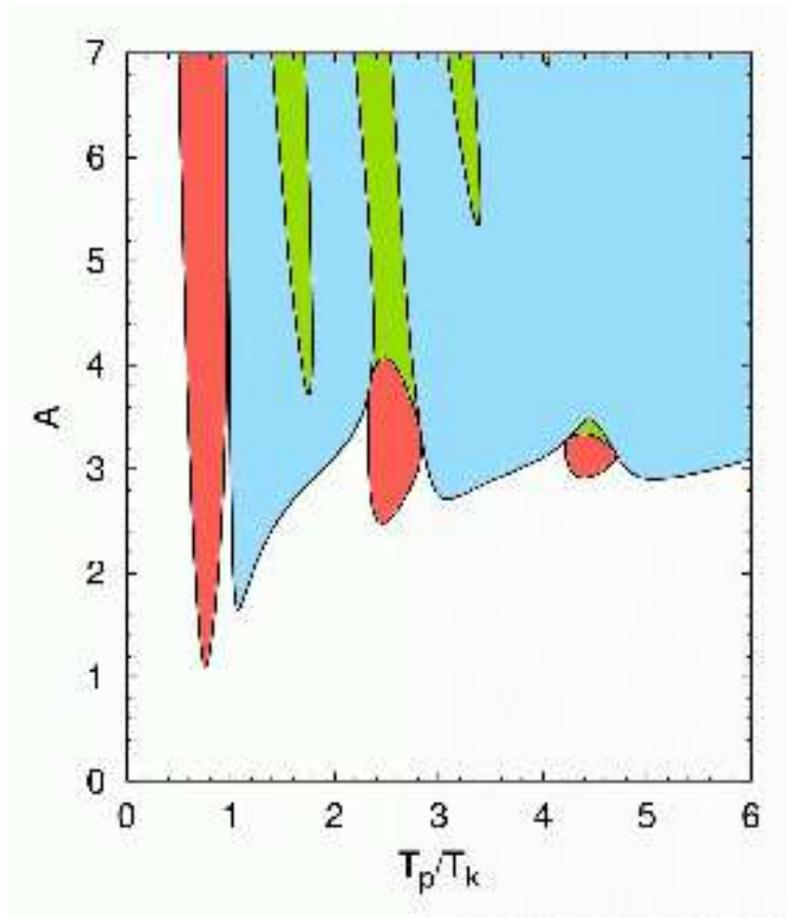}
\caption{Bifurcation diagram for the anti-phased dimer with the same
parameters as in Fig.~\ref{fig:phasehighgamma}}
\label{fig:phasehighgammadimer}
\end{center}
\end{figure}

Results for relatively large damping ($\gamma/\omega_0=0.4$) are
presented in the bifurcation diagram of
Fig.~\ref{fig:phasehighgammadimer}, which should be compared to
Fig.~\ref{fig:phasehighgamma}. The similarity between the two figures
is remarkable. Despite some extra green regions and the absence of the
yellow tongue, one notices that the principal resonance regions of the
dimer bifurcation diagram ($1/2 \lesssim T_p/T_k \lesssim 3/2$) fits
the same region in the mean field model almost exactly. The green
regions connected to pink regions in the mean field model are also
well mimicked by the dimer.

\begin{figure}[tb!]
\begin{center}
\includegraphics[height=4.8in]{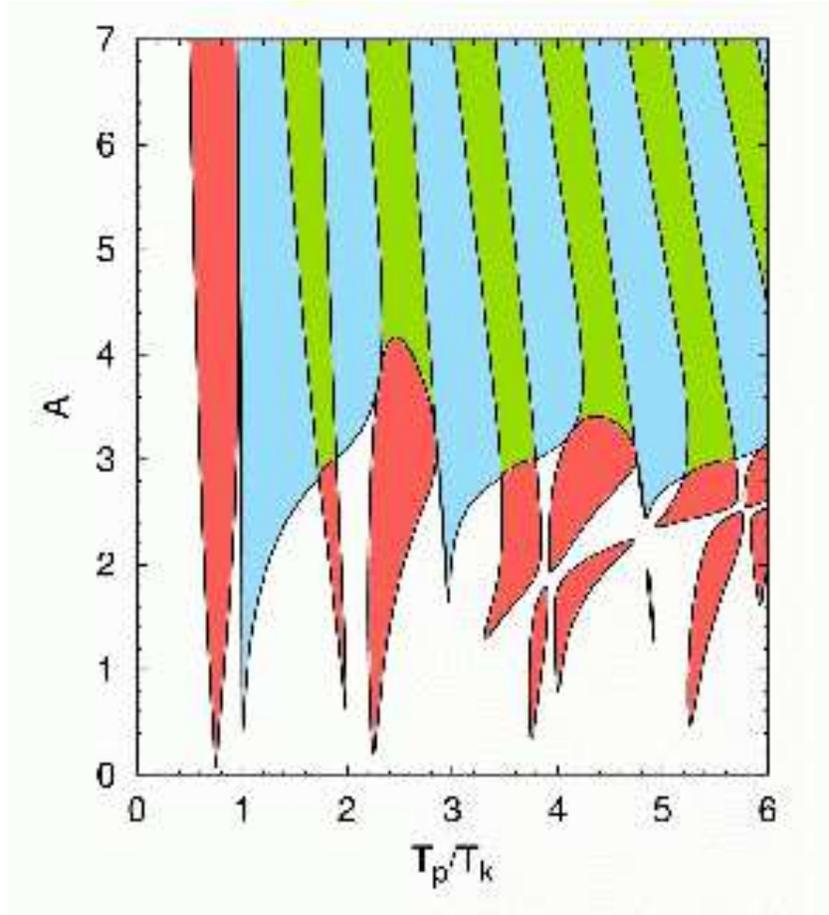}
\caption{Phase diagram for the anti-phased dimer with the same
parameters as in Fig.~\ref{fig:phaselowgamma}}
\label{fig:phaselowgammadimer}
\end{center}
\end{figure}

Figure~\ref{fig:phaselowgammadimer} illustrates the bifurcation
diagram for a small value of damping
($\gamma/\omega_{0}=0.01$). Comparison with
Fig.~\ref{fig:phaselowgamma} shows that although the agreement between
the two
models is not as good as for higher values of $\gamma/\omega_{0}$, 
the basic structure and similarities of the phase diagram and the
bifurcation diagram are nonetheless preserved. The main differences
are the complex pink patterns in region $T_p/T_k \gtrsim 3$ of the dimer. 
Also, as in
the high-gamma case, the dimer has larger green regions than the mean
field model, suggesting that the coupling in the latter plays a
stronger role in stabilizing the system. But in spite of these
differences and most importantly, the
principal resonance region ($1/2 \lesssim T_p/T_k \lesssim 3/2$) again
shows an almost perfect match.

In the dimer, competition between two kinds of synchronization plays a
key role in the destabilization of the system: on the one hand,
synchronization between each oscillator and its modulation; on the
other hand, synchronization between the two oscillators. This
competition is essentially governed by the values of $A$ and
$k$. Larger values of $A$ favor the former, while larger values of $k$
favor the latter. When the coupling is weak, the energy difference
between symmetric and antisymmetric modes is small and both can be
excited. In this case, the individual oscillators are nearly
independent and the stability diagram of the dimer is similar to that
of a single oscillator. As the coupling strength increases, the
energy of antisymmetric oscillations increases until eventually
only in-phase
oscillations are energetically accessible. This mutually synchronized
motion brings the system out of synchronization with the modulation.

\begin{figure}[tb!]
\begin{center}
\includegraphics[height=4.8in]{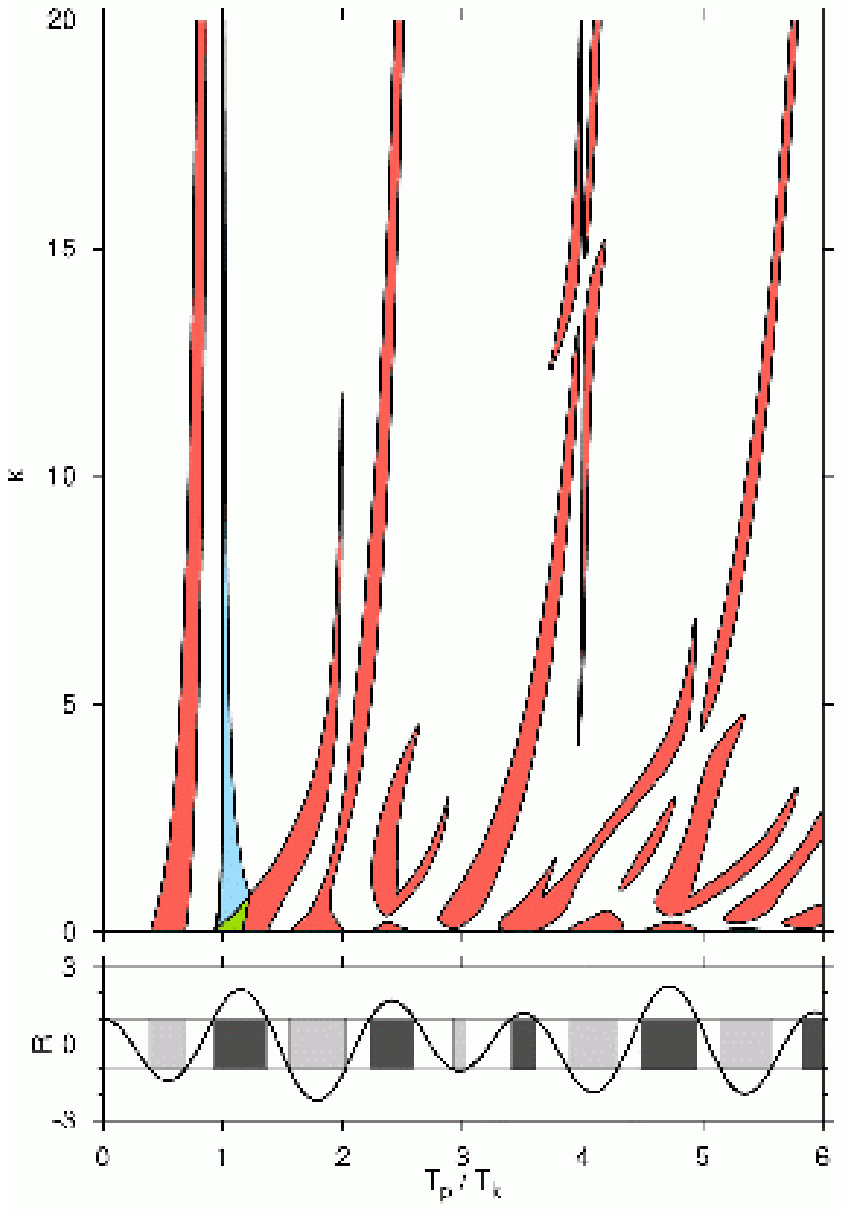}
\caption{Upper panel: Bifurcation diagram for the anti-phased dimer in the
$(T_p/T_k, k)$ plane for $A=0.9$, $\gamma=0.01$, and $\omega_0=0.4$, using
the same color convention as in Fig.~\ref{fig:phasehighgamma}.  Lower panel:
$R$ (Eq.~(\ref{eq:R})) as a function of $T_p/T_k$; light grey areas denote
$R<-1$ and dark grey areas $R>1$.}
\label{fig:krkplane}
\end{center}
\end{figure}

The stability diagram of the anti-phased dimer in the $(T_p/T_k,k)$
plane shown in Fig.~\ref{fig:krkplane} illustrates the above
story. First, consider $T_p/T_k \approx 0.5$, where a single
oscillator is in the main parametric instability region
($R<-1$, the first light grey region in the lower panel of
Fig.~\ref{fig:krkplane}). For
small $k$, the dimer is also unstable and still dominated by the
antisymmetric mode (even though the symmetric mode {\it cannot\/}
disappear, as mentioned above). However, as $k$ increases, excitation
of the antisymmetric mode becomes more difficult and the symmetric
mode becomes dominant, which stabilizes the system. When $0.5 \lesssim
T_p/T_k \lesssim 1$ a single oscillator is stable ($|R|<1$, between the
dark and light grey regions). In this
parameter region, individual oscillators do not have to be synchronous
to the modulation (see Fig.~\ref{fig:poles}). They are free to become
mutually synchronized and the system becomes unstable above a certain
coupling strength. Since the symmetric mode dominates, this
instability persists even for large $k$. Finally, when $T_p/T_k
\approx 1$, the single oscillator is again in an unstable region
($R>1$, the dark grey region). Although the situation is similar
to the first case, the
individual oscillators now have a zero-frequency mode which is not
subject to synchronization with the modulation. Therefore, the
zero-frequency mode of the two oscillators can be mutually
synchronized, which produces a monotonic growth of the mean. Again,
this instability does not have strong dependence on the coupling
strength and persists even for large $k$.

The good agreement between the dimer and mean field models is not 
merely a coincidence.  Consider one particular test oscillator in the
globally coupled system. Now define the group of its ``friends'' as
the set of all oscillators whose modulation phase lies within an
interval of $\pm \pi/2$ around its phase. The group of its ``enemies''
comprises the set of all other oscillators, whose average modulation
phase is opposite to that of the test oscillator. When the oscillators
are synchronized to the modulations, ``friends'' are also mutually
synchronized to one another, regardless of the coupling. When the
coupling increases, there is a competition between two kinds of
synchronization: synchronization between the ``enemies'' and
``friends'', and synchronization with their own modulations. This
situation is similar to the anti-phased dimer, which helps explain the
remarkable similarity between the instability diagrams of the dimer
and mean field models. Notice that this description becomes {\it
exact\/} in the quenched limit, where the ``friends'' and ``enemies''
can be represented respectively
by $\left<x\right>_+$ and $\left<x\right>_-$. From
this perspective, the choice $\theta=\pi$ for the dimer appears as a
natural one, being not only a particularly symmetric case in the
general dimer problem~\cite{mauro} but also the effective phase difference
between the two groups.

\section{Discussion and Conclusions}
\label{sec:conclusions}

We have investigated collective instabilities of an infinite set of globally
coupled linear oscillators driven by time-periodic piecewise linear
modulations with random
initial phases. These instabilities occur in certain parameter regimes (and
not in others), and we have produced phase diagrams as a function of system
parameters indicating detailed stability boundaries and types of
instabilities.  

Instabilities arise from phase synchronization (although not all
phase synchronization leads to instability), and there are two possible
competing synchronization mechanisms: synchronization of individual
oscillators with the external modulation (``modulation synchronization"),
and mutual synchronization between oscillators (``mutual synchronization"). 
In the absence of the external modulation, only mutual synchronization
is possible.  On the other hand, when the coupling is absent, a
single oscillator in parametric resonance condition ($|R|>1$) synchronizes
with the external modulation. Note that parametric instability
of a single oscillator requires modulation synchronization, and
in general {\em any} instability requires an external modulation to pump
energy into the system. 

In the presence of external modulation with random phases
{\em and} coupling, even in a parameter regime where either alone
would lead to synchronization, it is not possible for both types
of synchronization to occur simultaneously. Intuition might lead to the
conclusion that sufficiently strong coupling would favor mutual
synchronization thereby destroying modulation synchronization (and
hence stabilizing the system).  Intuition might also lead
one to conclude that weak coupling necessarily results in a dominance of
modulation synchronization and hence to instabilities of individual
oscillators not related to one another.  However, we have shown 
that this intuitive picture would be incomplete because it does
not account for the existence of {\em collective} parametric instabilities
due to the coupling and entirely absent in a single parametric
oscillator.  Here we discuss our results in terms of the two competing
synchronization mechanisms.

In a globally coupled model, the individual oscillators are coupled to
the mean $\left<x\right>$, and thus mutual synchronization can be thought of as
synchronization between the individuals and the mean. Note
that there is no mutual synchronization when $\left<x\right>=0$
(because if there were, the mean would not be zero).

Now suppose we are in the parameter regime $R<-1$ where individual
uncoupled oscillators are parametrically unstable.  When these oscillators
are coupled, one can imagine one of two possible scenarios.  If the coupling
leads to mutual synchronization, the individual oscillators can no longer be
synchronous with the external modulation and therefore the coupled system
has been stabilized by the coupling.  On the other hand, if the coupling
does not lead to mutual synchronization but instead there is modulation
synchronization, then the oscillators may be individually unstable but with
$\left<x\right>=0$.  Our results show that the second scenario is the
correct one for sufficiently small values of $k$, as shown
in the yellow regions of incoherent instability in
Fig.~\ref{fig:phaselowgamma}.  Modulation synchronization has ``won".  On
the other hand, there is a coupling energy cost to the lack of mutual
synchronization, which slows down the instability of individual oscillators
relative to their uncoupled amplitude growth.  For larger $k$ the first
scenario takes over, and the yellow region disappears above a certain value
of the coupling.

Next suppose that we are in the other parametric resonance regime, $R>1$.
The  situation is in some ways
similar to the previous case but there is a major difference:  
there is now a mode, the $j=0$ mode, whose frequency is zero and which
therefore (contrary to the other modes) need not (indeed cannot)
synchronize to the modulation.  The amplitude of this mode can grow
monotonically in either direction, and the coupling among oscillators leads
to a tendency for the zero-frequency mode of all the individual oscillators to
move in the same direction.  Thus while the growth rate of the $j \ne 0$
modes is reduced by coupling due to the lack of mutual synchronization,
that of the $j=0$ mode is enhanced because the coupling fosters 
mutual synchronization of this mode.

If the individual oscillators are not in regimes of parametric instability 
($|R|<1$),  there is no synchronization to the modulation and the
oscillators are free to synchronize with one another.  
Mutual synchronization is thus established and the mean $\left<x\right>$
becomes oscillatory with the same frequency as that of 
individual oscillators in the coupled system.  The oscillatory mean drives
the system into unstable states via the mean field coupling.

We have found that the instability bifurcation diagram for a
simple anti-phased dimer model reproduces the phase diagram for the mean
field system with surprising accuracy.

In the absence of external modulation, a dimer has two eigenmodes:
symmetric or mutually synchronized (lower energy), and antisymmetric
or mutually anti-synchronized (higher energy).  In the presence of
time-periodic piecewise linear modulations which are exactly
out of phase on the two coupled oscillators ($\theta=\pi$),
a competition between these two modes (which are no longer eigenmodes)
ensues. This competition is in many ways similar to
the competition between modulation synchronization and mutual
synchronization described for the globally coupled system, 
and here again it determines the stability of a dimer. When the
coupling is weak, the energy difference between symmetric and
antisymmetric modes is small and
both can be excited.  In this case, the individual oscillators are
nearly independent and the stability diagram of the dimer is similar
to that of a single oscillator.  The synchronization of each oscillator
with its modulation dominates the behavior, and instabilities thus
represent boundless excitation of the antisymmetric mode.
With increasing coupling the energy of the antisymmetric mode increases
until it is too high to be excited.  Only the symmetric mode 
can be excited, i.e., the oscillators become mutually synchronized.
The synchronization with the modulation is thus destroyed and the
associated parametric instability is suppressed.

Although the similarity between the dimer and globally coupled
models is remarkable, they also exhibit various important differences.  
In the dimer model, mutual synchronization involves only two oscillators.
On the other hand, in the global coupling model
an oscillator must be synchronous with essentially all others to create
collective motion.  Therefore, in the thermodynamic limit $N
\rightarrow \infty$, the collective instability in the globally
coupled system is a genuine phase transition, whereas the instabilities in
the dimer are simple bifurcations.  Nevertheless, the stability boundaries
and dynamics of the mean amplitudes in both cases show impressive
similarities.  

An interesting case that is in some sense ``in between" these two
and that promises interesting new features is that of
a one-dimensional chain of oscillators
with nearest-neighbor coupling.
When the phase of the modulations of the oscillators in the chain is
chosen at random, there is a significant chance that both neighbors
of any given oscillator have a phase ``similar" (suitably defined within
some range) to its phase.  In this case, the middle
oscillator can easily establish simultaneously both mutual
synchronization with its
neighbors and synchronization with the modulation.  Therefore, locally 
this oscillator could become unstable.  On the other hand, if the
neighbors of a given oscillator are modulated with phases opposite to
its own modulation phase, that oscillator may be stabilized. 
Therefore, the spatial pattern of the
modulation phase is expected to play an important role
and instability may become wavelength dependent, suggesting spatial
pattern formation.   Such patterns can of course not be observed in either a
dimer or a globally coupled model.  A detailed analysis of such systems will
be presented elsewhere~\cite{kawai}.

\section{Acknowledgments}
 
This work was supported in part by the National Science Foundation under
grant No. PHY-9970699.  Support of the Program on Inter-University
Attraction Poles of the Belgian Government and the F. W. O. Vlaanderen
(CVdB) is also gratefully acknowledged.

\appendix

\section{Time-evolution operator of a single parametric oscillator}
\label{app:GT}

We solve the equation of motion (\ref{eq:EOM_free}) 
using a standard Floquet method. The
damping term can be eliminated by introducing a new variable
$y$ defined as
\begin{equation}
x(t) = e^{-\gamma t/2} y(t).
\label{eq:x_to_y}
\end{equation} 
Equation~(\ref{eq:EOM_free}) becomes
\begin{equation}
\ddot{y}+ \omega^{2}(t) y = 0
\label{eq:EOM_y}
\end{equation}
with the time-dependent frequency
\begin{equation}
\omega^{2}(t) \equiv \omega_{0}^{2}[1 + \xi_\tau(t)] -
\gamma^{2}/4.
\end{equation}
The solution of the undamped frequency-modulated oscillator
(\ref{eq:EOM_y}) can be expressed in terms of the
time-evolution operator from $t=0$ to $t$, ${\sf g}_\tau (t)$,
as
\begin{equation}
\left(
\begin{array}{c}
y(t)\\
\dot{y}(t)
\end{array}
\right) =
{\sf g}_\tau (t) 
\left(
\begin{array}{c}
y(0)\\
\dot{y}(0)
\end{array}
\right)
\end{equation} 
For a piecewise constant modulation such as Eq.~(\ref{eq:perturbation}),
the explicit form of the time-evolution operator is known.
Using its periodicity and composition property we note that 
if $t=nT+u$ then
\begin{equation}
{\sf g}_\tau (t) = {\sf g}_\tau (nT+u) =
{\sf g}_\tau (u) {\sf g}_\tau (nT) =
{\sf g}_\tau (u) \left [{\sf g}_\tau (T) \right ]^{n} .
\label{eq:g_mult}
\end{equation}
It is thus sufficient to find ${\sf g}_\tau (t)$ for $t \in [0,T]$ 
 
When the phase is 
$\tau \in [0,T/2]$, the frequency varies as
\begin{equation}
\omega(t) =
\left \{
\begin{array}{ll}
\omega_+ & \quad \text{for} \quad t \in [0,\frac{\displaystyle T}
{\displaystyle 2}-\tau)  \\ [12pt]
\omega_- & \quad \text{for} \quad t \in [\frac{\displaystyle T}
{\displaystyle 2}-\tau,T-\tau)\\ [12pt]
\omega_+ & \quad \text{for} \quad t \in [T-\tau,T]
\end{array}
\right. 
\label{eq:omega1}
\end{equation}
while for $\tau \in [T/2,T]$,
\begin{equation}
\omega(t) =
\left \{
\begin{array}{ll}
\omega_- & \quad \text{for} \quad t \in [0,T-\tau)  \\ [12pt]
\omega_+ & \quad \text{for} \quad t \in [T-\tau,
\frac{\displaystyle 3T}{\displaystyle 2}-\tau)\\ [12pt]
\omega_- & \quad \text{for} \quad t \in [\frac{\displaystyle 3T}
{\displaystyle 2}-\tau,T]
\end{array}
\right. 
\label{eq:omega2}
\end{equation}
where the $\omega_\pm$ are defined in Eq.~(\ref{eq:omega_pm}).
During each constant frequency time window the system evolves 
according to the propagator of a simple harmonic oscillator
of the appropriate frequency.  This propagator is well known:
\begin{equation}
{\sf g}_\pm (t,t') = 
\left (
\begin{array}{cc}
\cos [\omega_\pm (t-t')] & \frac{\displaystyle 1}
{\displaystyle \omega_\pm} \sin [\omega_\pm (t-t')]
\\ [12pt]
-\omega_\pm \sin [\omega_\pm (t-t')] & \cos [\omega_\pm (t-t')]
\end{array}
\right ) .
\label{eq:free_g}
\end{equation}
The full operator ${\sf g}_\tau (t)$ 
can be expressed as products of the $g_\pm$.
For the case (\ref{eq:omega1}),
\begin{equation}
{\sf g}_\tau(t) =
\left \{
\begin{array}{ll}
{\sf g}_+ (t,0) & \quad \text{for} \quad t \in [0,\frac{\displaystyle T}
{\displaystyle 2}-\tau)  \\ [12pt]
{\sf g}_- \left(t,\frac{\displaystyle T}{\displaystyle 2}
-\tau\right) {\sf g}_+ \left(\frac{\displaystyle T}{\displaystyle 2}
-\tau,0\right) & \quad \text{for} \quad
t \in [\frac{\displaystyle T}{\displaystyle 2} -\tau,T-\tau)\\ [12pt]
{\sf g}_+ (t,T-\tau) {\sf g}_- \left(T-\tau,\frac{\displaystyle T}
{\displaystyle 2}-\tau\right) {\sf g}_+ \left(\frac{\displaystyle T}
{\displaystyle 2}-\tau,0\right) & \quad \text{for} \quad t \in [T-\tau,T]
\end{array}
\right.
\label{eq:gt1}
\end{equation}
and for the case (\ref{eq:omega2}),
\begin{equation}
{\sf g}_\tau(t) =
\left \{
\begin{array}{ll}
{\sf g}_- (t,0) & \quad \text{for} \quad t \in [0,T-\tau)  \\ [12pt]
{\sf g}_+ (t,T-\tau) {\sf g}_- (T-\tau,0) & \quad \text{for} \quad
t \in [T-\tau,\frac{\displaystyle 3T}{\displaystyle 2}-\tau)\\  [12pt]
{\sf g}_- \left(t,\frac{\displaystyle 3T}{\displaystyle 2}-\tau\right)
{\sf g}_+ \left(\frac{\displaystyle 3T}{\displaystyle
2}-\tau,T-\tau\right)
{\sf g}_- (T-\tau,0) & \quad \text{for} \quad t
\in [\frac{\displaystyle 3T}{\displaystyle 2}-\tau,T].
\end{array}
\right.
\label{eq:gt2}
\end{equation}
These expressions can be further simplifed by using time-translation
symmetry, ${\sf g}_\pm (t+u,t'+u) =
{\sf g}_\pm (t,t') = {\sf g}_\pm (t-t',0)$.   We can thus simplify our
notation and set ${\sf g}_\pm (t,0) \equiv {\sf g}_\pm (t)$ with no loss
of information.  

In particular, for $t=T$
\begin{equation}
{\sf g}_\tau (T) = {\sf S}_\tau {\sf g}_0 (T) {\sf S}^{-1}_\tau
\label{eq:gT}
\end{equation}
where
\begin{equation}
\begin{array}{lll}
{\sf g}_0 (T) \equiv {\sf g}_- \left(\frac{\displaystyle T}
{\displaystyle 2}\right) {\sf g}_+ \left(\frac{\displaystyle T}
{\displaystyle 2}\right), \quad
& {\sf S}_\tau \equiv {\sf g}_+(\tau), \quad
& \text{for} \quad \tau \in [0,\frac{\displaystyle T}
{\displaystyle 2}] \\ [12pt]
{\sf g}_0 (T) \equiv {\sf g}_+ \left(\frac{\displaystyle T}
{\displaystyle 2}\right) {\sf g}_- \left(\frac{\displaystyle T}
{\displaystyle 2}\right), \quad
& {\sf S}_\tau \equiv {\sf g}_-\left(\tau-\frac{\displaystyle T}
{\displaystyle 2}\right), \quad
& \text{for} \quad \tau \in [\frac{\displaystyle T}{\displaystyle 2},T].
\end{array}
\label{eq:g0}
\end{equation}
Since only ${\sf S}_\tau$ depends on $\tau$, the trace, determinant, and
eigenvalues of ${\sf g}_\tau (T)$ do not depend on the random phase. 

Transforming back to the original variables, 
we finally obtain the time-evolution operator for $x(t)$ and $\dot{x}(t)$,
\begin{equation}
{\sf G}_\tau (t) = e^{-\gamma t/2} {\sf g}_\tau (t) .
\label{eq:GT}
\end{equation}

Next we derive an explicit expresion for the Laplace transform of the
evolution operator with the help of Eq. (\ref{eq:g_mult}): 
\begin{eqnarray}
\tilde{\sf G}_\tau(s) &=&  
\int_0^\infty e^{-st} {\sf G}_\tau(t) dt
\nonumber\\ [12pt]
&=& \sum_{m=0}^{\infty} 
\left [ e^{-sT} {\sf G}_\tau(T) \right ] 
^m \int_0^T e^{-st} {\sf G}_\tau(t) dt
\nonumber\\ [12pt]
&=&
\left (I - e^{-sT}{\sf G}_\tau(T) \right )^{-1} 
\int_0^T e^{-st} {\sf G}_\tau(t) dt 
\nonumber \\ [12pt]
&=& \frac{\frac{1}{2} e^{-\gamma T/2}
\left[e^{(s+\gamma)T} I  - {\sf G}_\tau(T) \right]
\int_0^{T} e^{-st} {\sf G}_\tau(t) dt }
{\cosh[\left(s+\frac{\displaystyle \gamma}{\displaystyle 2}\right)T]- R }             
\label{eq:GsApp}
\end{eqnarray}
where $I$ is an identity matrix and
\begin{eqnarray}
R &=& \frac{1}{2}e^{\gamma T/2}  \text{Tr } {\sf G}_0 (T) \nonumber \\
[12pt]
&=& \frac{1}{2} \text{Tr } {\sf g}_0(T) \nonumber \\
[12pt]
&=&  \cos(\frac{\omega_{+}T}{2}) \cos(\frac{\omega_{-}T}{2}) 
     -\frac{\omega_{+}^{2} + \omega_{-}^{2}}{2\omega_{+} \omega_{-}} 
      \sin(\frac{\omega_{+}T}{2})  \sin(\frac{\omega_{-}T}{2}) 
\end{eqnarray}
The geometric series in Eq. (\ref{eq:GsApp}) converges 
only when  
\begin{equation}
\|e^{-sT} {\sf G}_0 (T) \| < 1.
\label{eq:cond_gsum}
\end{equation}

\clearpage

\clearpage

\begin{thebibliography}{99}
\bibitem{par} Such parametric oscillators are encountered in a
variety of
contexts that include elementary particles, astrophysics, fluid
mechanics, magnetism, plasma physics, electrical and electronic
networks, optical systems and lasers, applied mechanics, and biophysics,
cf. bibliography in \cite{old} and \cite{mauro}.
\end{thebibliography}
\end{document}